\documentclass[sn-mathphys,Numbered]{sn-jnl}

\usepackage{graphicx}%
\usepackage{multirow}%
\usepackage{amsmath,amssymb,amsfonts}%
\usepackage{amsthm}%
\usepackage{mathrsfs}%
\usepackage[title]{appendix}%
\usepackage{xcolor}%
\usepackage{textcomp}%
\usepackage{manyfoot}%
\usepackage{booktabs}%
\usepackage{algorithm}%
\usepackage{algorithmicx}%
\usepackage{algpseudocode}%
\usepackage{listings}%
\usepackage{csquotes}
\usepackage{amsmath}
\usepackage{pdflscape}
\usepackage{breqn}
\usepackage{subcaption}
\usepackage{tikz}
\usepackage{csquotes}
\usepackage{tipa}

\usepackage{graphicx}
\usepackage{epstopdf, epsfig}
\usepackage{array}
\usepackage{marginnote}
\usepackage{hyperref}

\hypersetup{
    colorlinks=true,    
    linkcolor=black,    
    citecolor=black,    
    urlcolor=blue       
}

\theoremstyle{thmstyleone}%
\theoremstyle{thmstyletwo}%

\theoremstyle{thmstylethree}%

\begin{document}

\title[Article Title]{Experimentally informed, linear mean-field modelling of circular cylinder aeroacoustics}


\author[1]{\fnm{Robin} \sur{Prinja}}\email{robin.prinja@univ-poitiers.fr}

\author[1]{\fnm{Peter} \sur{Jordan}}\email{peter.jordan@univ-poitiers.fr}

\author*[1]{\fnm{Florent} \sur{Margnat}}\email{florent.margnat@univ-poitiers.fr}

\affil*[1]{\orgdiv{D{\'e}partement Fluides Thermique et Combustion}, \orgname{Institut Pprime, Universit{\'e} de Poitiers - CNRS - ENSMA}, \orgaddress{\city{Poitiers}, \postcode{86000},  \country{France}}}


\abstract{A noise modelling approach is proposed for bluff body wakes such as flow over a cylinder, 
where the primary noise source comprises large-scale coherent structures such as the vortex shedding flow feature.
This phenomenon leads to Aeolian tones in the far-field, and is inherent in wake flows across a range of Reynolds numbers (Re), from low-Re to high-Re turbulent flows.
The approach employs linear global stability analysis on the time-averaged mean flow, with amplitude calibration through two-point statistics, and far-field noise calculations from the global mode fluctuations by Curle's analogy.
The overall approach is tested for flow over a cylinder at Reynolds numbers Re=150 and 13300. 
For Re=150 flow, noise directivity calculations from the present approach agree with direct far-field computations.
For Re=13300 flow, the mean flow is obtained by particle image velocimetry (PIV). The linear global mode for spanwise-homogeneous-type fluctuations is obtained at the main, lift fluctuation frequency. Calibration of this global mode involves Time-Resolved PIV data in the streamwise-spanwise plane, which is Fourier transformed in frequency-spanwise wavenumber space. The noise calculations for this global mode are then found to be less than $1$ dB off from the microphone measurements.
}






\maketitle

\clearpage

\clearpage
\section{Introduction} \label{sec:intro}


Bluff body noise is one of the most important components of airframe noise during the landing phase of commercial aircraft. Landing gear and high lift devices such as trailing edge
flaps, leading-edge slats are the primary contributors to the airframe noise \citep{li2013control,dobrzynski1997full}. 
Similar noise sources can be found in other
moving bodies, for instance, a car’s rear-view mirror and an electric train’s pantograph.

\citet{strouhal1878besondere} characterized the tones generated by a flow over a circular cylinder, now known as Aeolian tones.
For a flow, with free stream velocity as $U_{\infty}$, over a bluff-body of characteristic length $d$, the scaling law for the Aeolian tone frequency was found as 
\begin{equation}
\text{St}=\frac{fd}{U_{\infty}},
\end{equation}
where St is the Strouhal number.
Accompanied by background turbulence, the flow exhibits the von K\'{a}rm\'{a}n wake \citep{von1911mechanismus}, an organized vortex-shedding feature which consists of a periodic train of counter-rotating vortices shed in the wake of bluff bodies, which leads to
Aeolian tones in the acoustic field \cite{phillips1956intnesity}.
As the flow passes over the bluff body, it creates boundary layers on both the upper and bottom surfaces of the cylinder. These boundary layers separate, resulting in the formation of two shear layers whose unstable character leads to the appearance of the von K\'{a}rm\'{a}n wake.

Flow over a cylinder exhibits distinct characteristics depending on the Reynolds number  $\mathrm{Re}=U_{\infty}d/\nu$ where $\nu$ is the kinematic viscosity of the fluid. For Re values greater than 47, the flow displays two-dimensional vortex shedding; for Re values greater than 200, three-dimensionality becomes apparent. At Re values greater than 400, the wake becomes turbulent, and for Re values exceeding $10^6$, the boundary layers on the cylinder surface become turbulent \citep{jackson1987finite,zdravkovich1997flow}.
A comprehensive discussion of bluff-body wake flows can be found in the reviews by \citet{williamson1996vortex}, \citet{rajagopalan2005flow}, and \citet{derakhshandeh2019review}.

Prediction of bluff-body noise can be achieved by solving the Navier-Stokes equations directly in the unsteady, compressible regime. This returns both the turbulent and acoustic field simultaneously  \cite{inoue2002sound,muller2008high}) or in conjunction with acoustic analogies such as Lighthill's  \citep{lighthill1952sound} and its extensions \cite{ffowcs1969sound, curle1955influence}. 
The computational expense of such models for high-Re flows and complicated geometries, however, remains substantial primarily due to the necessity of resolving a broad spectrum of turbulent scales over a 3D computational domain and a significant physical time. 

Simplified physics-based noise models, on the other hand, are generally desired in the design of low-noise airframe structures. 
These models offer a crucial physical understanding of sound-generation mechanisms, making them invaluable for the initial design phases of quieter airframe structures. \citet{phillips1956intnesity} derived the following law for Aeolian tones:
\begin{equation}
\displaystyle p^2_{\mathrm{rms}}(r,\theta) = 0.037 ~ \frac{\sin^2\theta}{r^2}~\frac{\rho^2 U_\infty^6}{c_0^2} ~\mathrm{St}^2 ~L d  ,
\label{eq:phillips_model} \end{equation}
where $p_{\mathrm{rms}}$ is the root-mean-square (rms) pressure fluctuations at a far-field distant $r$ from the cylinder center and at an angle $\theta$ with the free stream flow, $L$ is the cylinder length, $\rho$ is the fluid density, and $c_0$ is the free-stream sound speed. The constant was obtained empirically in order to match with measurements in the range $360< \text{Re}<30000$. It can be made explicite in the far-field for low Mach number, assuming the cylinder as acoustically compact in the transverse plane. Derived from Curle's analogy \citep{curle1955influence}, the refined formula is \cite{phillips1956intnesity,fujita2010chars},
\begin{equation}\label{eq:fujita_model}
\displaystyle p^2_{\mathrm{rms}}(r,\theta) = \frac{\sin^2\theta}{r^2}~\frac{\rho^2 U_\infty^6}{c_0^2} ~\mathrm{St}^2 ~ \frac{L L_c C_l^2}{16}  ,
\end{equation}
where $C_l$ is sectional fluctuating (rms) lift coefficient on the cylinder and $L_c$ is the spanwise correlation length or spanwise coherence length, representing length scale of the spanwise decay of two-point correlation function~\cite{phillips1956intnesity} or coherence function \citep{Kato93,Fujita1998,margnat2023cylinder} respectively. 
These lengths are essential components of various current bluff body noise models \cite{casalino2003prediction, fujita2010chars, kato1993numerical, seo2007aerodynamic, Doolan2010}, referred to as $L L_c$ models. These latter rely on the assumptions that: (i) The spanwise distribution of the vortex-shedding phase is random; (ii) a single parameter, $L_c$, can comprehensively represent the spanwise phase of the acoustic source. 
Another limitation of such models is that it is not clear which part of the spanwise dynamics dominates as acoustic source.

A rather different approach considers coherent structures in the flow, defined as organized and persistent patterns that exist in turbulent flows, and that exhibit spatial and temporal scales significantly larger than the integral turbulence scales. Their existence in several turbulent shear flows has been demonstrated in cylinder wakes \citep{townsend1956structure, grant1958large, williamson1996vortex}. 


Coherent structures can be educed from time-resolved flow field data \citep{lumley1967structure, lumley1981coherent} by means of various data-processing procedures \citep{taira2017modal, taira2020modal} 
such as proper orthogonal decomposition (POD) \citep{lumley1967structure, towne2018spectral}, 
dynamic mode decomposition (DMD) \citep{rowley2005model},
linear stochastic estimation \citep{adrian1977ontherole}, 
and wavelet analysis \citep{mallat1989theory,farge1992wavelet, camussi2021application}.
Also, Fourier decomposition can be used for the homogeneous directions, if they exist, in the data.
These techniques enable the sorting of turbulent flows into modes ranked based on various criteria, such as fluctuation energy content or acoustic efficiency, potentially resulting in a low-order system in some cases.

For axisymmetric jets, it has been shown that, when decomposed in terms of azimuthal Fourier modes, the lowest order modes are the most efficient acoustic sources \citep{ jordan2013wave, cavalieri2012axisymmetric, cavalieri2013wavepackets}.
This is in fact a property of the Green function that describes the problem, resulting in the dominance of low azimuthal modes in sound radiations despite their low fluctuation energy in comparison to the energy-containing flow scales: the first three lowest-order azimuthal Fourier modes contribute less than $10\%$  to the overall fluctuation energy \citep{cavalieri2012axisymmetric, jordan2013wave, jaunet2017two}. For flow over a forward-facing thick plate, 
\citet{debesse2016comparison} performed an LES and analysed the resulting database with means of Fourier analysis, POD and DMD in order to study the dominant noise producing flow motions. 
The spanwise structure of the dispersion relation was used to establish that, in terms of acoustic efficiency, only the spanwise homogeneous mode i.e. $k=0$ Fourier mode can be efficient in driving propagative pressure fluctuations.

For the flow over a spanwise-homogeneous cylinder, the spanwise Fourier decomposition of the flow field  has not been explored yet. 
In the context of vortex-shedding fluctuations, the source of Aeolian tones, if a low-order system representation of the dominant acoustic sources is found, formulation of noise models which are low-order, simplified and physics-based could be done.
Advantage of such models over $L L_c$ models is that they allow us to separately analyze each spanwise Fourier mode's contribution to far-field noise, hence providing a clearer understanding of the spanwise organisation of the dominant acoustic sources.

For such situations, where a low-order respresentation of the acoustically dominant coherent structures is present, 
noise models based on linear stability analysis around mean-flow emerge as a good candidate.
It has been found that many important characteristics of coherent structures, such as their spatial structures, and phase speeds, can be described using linearised flow equations, where linearization is performed about the mean flow in flows such as jets \citep{crighton1976stability, crow1971orderly, gudmundsson2011instability, suzuki2006instability, cavalieri2013wavepackets}, boundary layers \citep{schoppa2002coherent}, and airfoils \citep{de2014global, yeh2019resolvent, symon2019tale, abreu2021spanwise, demange2023resolvent}. 
However, due to the linear nature of the approach, a calibration step for the amplitudes and phases for linear instability modes is required before incorporating them as acoustic sources in the noise models.

For low-Re flows (Re$\leq 150$) over a circular cylinder, linear stability analyses about mean flow fields have already proven useful in terms of modelling the phenomenon of vortex shedding \citep{huerre1990local, pier2002frequency, chomaz2005global, giannetti2007structural, sipp2007global, fani2018computation,  barkley2006linear, meliga2012sensitivity}.
\citet{barkley2006linear} was the first to conduct a comprehensive 2D linear stability analysis of the mean flow in the cylinder wake across Reynolds numbers ranging from 46 to 180, providing eigenfrequencies, growth rates and eigenstructures for the general 2D perturbations.

It has been shown that, beyond the threshold Reynolds number ($\text{Re}>47$), the mean flow, rather than the base flow which is the fixed-point solution of the Navier-Stokes system, yields the most accurate profile for modelling the vortex shedding phenomenon
\cite{pier2002frequency, barkley2006linear, mittal2008global}.
The inaccuracy in the frequency prediction from base-flow stability analysis is due to the invalidity of the linear approximation as the small unstable perturbations keep on growing far beyond the linear approximations.
 Initiating from the base flow, studies \cite{maurel1995mean, zielinska1997strongly, noack2003hierarchy} demonstrate modifications to the mean flow due to the oscillating wake and subsequent nonlinear saturation via interaction with the mean flow. The model proposed by \citet{noack2003hierarchy} suggests that the amplitude of the oscillating wake saturates precisely when the mean flow is marginally stable—a concept reminiscent of the marginal stability criterion proposed by \citet{malkus1956outline} for fully developed turbulent flows. 
 
\citet{triantafyllou1986formation} found that the formation of the vortex street is due to an absolute instability in the wake immediately behind the cylinder, and the appearance of global instability can be seen as the development of a region of absolute instability. 
Due to this absolute instability, any initial disturbance grows at any fixed location and after nonlinearities have limited the growth of the disturbance, a self-sustained oscillation of the wake is established. 
For more in-depth reviews on global instability, readers are referred to works by \citet{chomaz2005global, theofilis2011global, sipp2010dynamics, taira2017modal, taira2020modal}.

 \citet{mantivc2014self, mantivc2015self} developed a self-consistent model based on global modes for cylinder flow, accurately predicting the frequency and spatial structure of the vortex shedding mode for Reynolds numbers up to 110. \citet{fani2018computation} extended this model to the compressible linearized Navier-Stokes equations, providing detailed instability information along with the acoustic field. These models were limited to Reynolds numbers below 150, where the flow remains laminar and two-dimensional, and required a starting base flow.


As the Reynolds number surpasses 400, wake flows behind cylinders transition to turbulence, marked by three-dimensional and turbulent characteristics. The applicability and effectiveness of linear mean-flow-based global stability noise models at such high Reynolds numbers are not yet fully explored, forming the focus of the current investigation.
It also remains to be seen whether, in cylinder flows, coherent structures associated with low-order spanwise wavenumbers serve as the primary sources for Aeolian tones, analogous to the scenario observed in axisymmetric jets.

In the present work, we present a physics-based simplified noise model for turbulent bluff body flows which is designed to represent coherent structures through a mean-flow-based 2D linear global stability analysis.
In the present model, coherent structures corresponding to different spanwise Fourier modes can be included individually as acoustic sources. With such an approach, we aim to improve the physical understanding of the flow in terms of its dominant acoustic source structure.

The paper is structured as follows: Section \ref{sec:Methodology} outlines the methodology of the noise model, with Section \ref{ssec:method_gsa} detailing the global stability analysis framework, and Section \ref{ssec:method_curle} presenting Curle's analogy for Aeolian tone. In Section \ref{sec:results_re150}, we apply the noise model to a Re=150 flow over a circular cylinder with validation \textit{versus} direct noise computation. 
Section \ref{sec:results_re13300} presents the application of the noise  model to a Re=13300 flow over a circular cylinder using experimental data for mean flow estimation and mode amplitude calibration. 
Finally, Section \ref{sec:conclusion} concludes the paper.


\section{Methodology} \label{sec:Methodology}
In the present work, we propose a model for the aeroacoustics of bluff body wakes, that exhibit tonal noise. The model is based on linear global stability analysis of the time-averaged of the flow around the bluff body, in which the leading global modes are considered to represent the acoustically important, coherent structures of the wake. 

\begin{figure}[t!]
     \centering
\begin{subfigure}{0.35\textwidth}
	\includegraphics[width=1\textwidth]{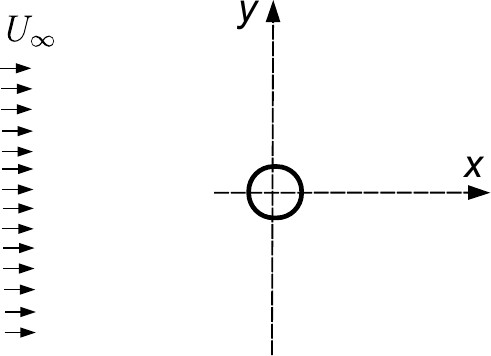}
        \caption{Front view.}
\end{subfigure}
     \hfill
\begin{subfigure}{0.35\textwidth}
	 \includegraphics[width=1\textwidth]{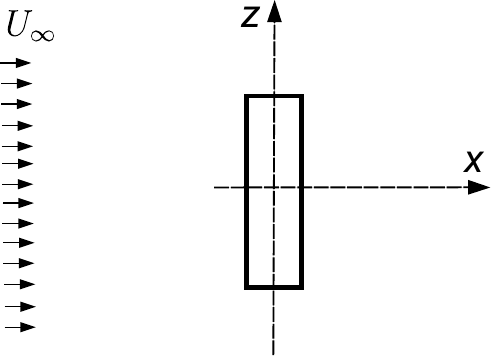}
        \caption{Bottom view.}
\end{subfigure}
        \caption{Flow over a spanwise-homogeneous cylinder: Coordinate system sketch. The origin is fixed at the cylinder's center at its mid-span. Free-stream flow, with velocity $U_{\infty}$, is along the $x-$axis. The transverse axis is along the $y-$axis. Cylinder's span is along the $z-$axis.}
        \label{fig:sketch}
\end{figure}

The application of this model in flow over a spanwise-homogeneous circular cylinder is demonstrated for two cases: Re=150 and Re=13300, in Section \ref{sec:results_re150} and Section \ref{sec:results_re13300} respectively. 
Re=150 flow is examined numerically by using data from a Direct Numerical Simulation (DNS), as an initial application of the proposed model. This serves as a numerical test of the methodology before tackling the more challenging experimental case of Re=13300 flow.
The coordinate system is depicted in \autoref{fig:sketch}.\\

The multi-step framework for the proposed models involves:
\begin{enumerate}
\item Obtain the time-averaged mean flow, including mean streamwise velocity ($\overline{U}$) and mean transverse velocity ($\overline{V}$). For the present applications, the time-averaged mean flow in the 2D plane normal to the cylinder axis, i.e. in the XY plane, is required. This is obtained here by DNS and PIV measurements for Re=150 and Re=13300 flows respectively.

\item Conduct linear global stability analysis on the time-averaged mean flow to extract phase speeds, fluctuation frequencies, and spatial structures of dominant global modes. This step is detailed in Section \ref{ssec:method_gsa}. Global stability analysis is done in the present work via Arnoldi iteration-based Matrix-free method \citep{arnoldi1951principle, lehoucq1998arpack}, by employing the Nektar++ code package \citep{cantwell2015nektar++}. 

\item Calibrate amplitudes of the dominant global modes. While the linear global modes may accurately capture most of the important features of the coherent structures, their amplitudes require calibration due to the linear nature of the approach. For the Re=13300 flow, this calibration is accomplished through two-point velocity measurements in the spanwise direction via TR PIV. These measurements are decomposed into spanwise Fourier modes to extract the individual amplitudes of the global modes corresponding to these spanwise Fourier modes. This is detailed in Section \ref{sssec:re13300_calibration}. For the Re=150 flow over a circular cylinder, as it is a two-dimensional flow, the global mode is calibrated by the one-point DNS recordings.

\item Utilize Curle's analogy to predict far-field noise associated with the calibrated global mode, considering it as the principal acoustic source, as presented in Section \ref{ssec:method_curle}.

\end{enumerate}

\subsection{Global stability analysis} \label{ssec:method_gsa}

The governing incompressible Navier-Stokes system in non-dimensional form can be written as:
\begin{subequations}\label{eqn:ins}
\begin{equation}\label{eqn:ins_cont}
\nabla \cdot {\mathbf{U}}=0,
\end{equation}
\begin{equation}\label{eqn:ins_mom}
\frac{\partial \mathbf{U}}{\partial t}+({\mathbf{U} \cdot \nabla }){\mathbf{{U}}}=-\nabla P +\frac{1}{Re} \Delta{\mathbf{U}}.
\end{equation}
\end{subequations}
where $\mathbf{U}$ is the velocity field vector, $P$ is the pressure field.
Reynolds decomposition of the flow variables into their time-averaged and fluctuating parts  ($\mathbf{U}=\mathbf{\overline{U}}+\mathbf{u}$ and $P={\overline{P}}+{p}$), gives a system of equations for the fluctuation field as
\begin{subequations}\label{eqn:tempo1}
\begin{equation}
\nabla \cdot {\mathbf{u}}=0,
\end{equation}
\begin{equation}
\frac{\partial \mathbf{u}}{\partial t}+
({\mathbf{\overline{U}} \cdot \nabla) }{\mathbf{u}}+
({\mathbf{u} \cdot \nabla )}{\mathbf{\overline{U}}}+
({\mathbf{u} \cdot \nabla) }{\mathbf{u}}
-\overline{({\mathbf{u} \cdot \nabla) }{\mathbf{u}}}
= -\nabla p +\frac{1}{Re}\Delta{\mathbf{u}},
\end{equation}
\end{subequations}
The non-linear term, $f=({\mathbf{u} \cdot \nabla) }{\mathbf{u}} 
-\overline{({\mathbf{u} \cdot \nabla) }{\mathbf{u}}}$, can be neglected to arrive at 
Linearized Incompressible Navier-Stokes equations,
\begin{subequations}\label{eqn:lins}
\begin{equation}
\nabla \cdot {\mathbf{u}}=0,
\end{equation}
\begin{equation}
\frac{\partial \mathbf{u}}{\partial t}+({\mathbf{\overline{U}} \cdot \nabla }){\mathbf{u}}
+({\mathbf{u} \cdot \nabla }){\mathbf{\overline{U}}}
=-\nabla p +\frac{1}{Re} \Delta{\mathbf{u}},
\end{equation}
\end{subequations}
System \eqref{eqn:lins} can be written in matrix form as,
\begin{equation}\label{eq:lns_matrix}
\mathcal{B} \dfrac{\partial \textbf{q}}{\partial t} = \mathcal{A} \textbf{q}
\end{equation}
where $\textbf{q}$ is the fluctuation field  expressed in vector form as
\begin{equation}
\textbf{q}=\begin{bmatrix}
u  \\
v  \\
w  \\
p
\end{bmatrix}=\begin{bmatrix}
\textbf{u}  \\
p
\end{bmatrix},
\end{equation}
$\mathcal{B}$ is a mass matrix given by 
\begin{equation}
\mathcal{B} =\begin{bmatrix}
1 & 0 & 0 & 0  \\
0 & 1 & 0  & 0  \\
0 & 0 & 1  & 0 \\
0 & 0 & 0 & 0  
\end{bmatrix}
\end{equation}
and $\mathcal{A}$ contains the information about the Navier-Stokes system, mean-flow information, Jacobians, and spatial derivates.

In the case of a spanwise-homogeneous bluff body, there exist two homogeneous directions, one in time and the other in its spanwise direction, allowing us to express disturbance around the mean flow in the following normal-mode form, 
\begin{equation}\label{eq:normal_zt}
\mathbf{q}(x,y,z,t)=\hat{\mathbf{q}}(x,y,z)e^{i\omega t}=\widetilde{\mathbf{q}}(x,y)e^{i k z}e^{i\omega t}.
\end{equation}
where $\omega$ is the complex frequency, $k$ is the spanwise wavenumber and $\widetilde{\mathbf{q}}(x,y)$ are the eigenfunctions (the spatial distribution of the fluctuation mode in the $x-y$ domain) for the mode for the corresponding $k-\omega$ wavenumber-frequency set.
Using the normal mode ansatz \eqref{eq:normal_zt} in the LNS \eqref{eq:lns_matrix} gives the 3D eigenvalue problem as 
\begin{equation}\label{eq:simple_eigen2}
i \omega  \mathcal{B} \widetilde{\mathbf{q}}(x,y) =
\mathcal{A}_k
 \widetilde{\mathbf{q}}(x,y) ,
\end{equation}
which can be solved as an eigenvalue problem for each spanwise wavenumber, $k$, individually, where we look for $\omega$ as eigenvalues (temporal) and $\widetilde{\mathbf{q}}(x,y)$ as eigenfunctions of the system. The final expression for $\mathcal{A}_k$ is:
\begin{equation}
 	\renewcommand{\arraystretch}{3}
\mathcal{A}_k =
{\footnotesize
\begin{bmatrix}
-U\dfrac{\partial}{\partial x}-V\dfrac{\partial}{\partial y}-\dfrac{\partial U}{\partial x} 
+ \dfrac{1}{Re} \Delta
& -\dfrac{\partial U}{\partial y} & 0 & -\dfrac{\partial }{\partial x} \\
-\dfrac{\partial V}{\partial x} & -U\dfrac{\partial}{\partial x}-V\dfrac{\partial}{\partial y}-\dfrac{\partial V}{\partial y} 
+ \dfrac{1}{Re}\Delta
&  0 & -\dfrac{\partial }{\partial y} \\
0 & 0 & 
-U\dfrac{\partial}{\partial x}-V\dfrac{\partial}{\partial y}+ \dfrac{1}{Re}\Delta
& -ik  \\
\dfrac{\partial}{\partial x} & \dfrac{\partial}{\partial y} & ik & 0 
\end{bmatrix}.
}
\end{equation}
where $\Delta=\left[\dfrac{\partial^2}{\partial x^2}+\dfrac{\partial^2}{\partial y^2}+\dfrac{\partial^2}{\partial z^2}\right]$.

As in the present case, the linear analysis is performed about the mean flow around a cylinder, we expect eigenvalues to be either stable or marginally stable. 
Solving the eigenvalue problem \eqref{eq:simple_eigen2} for 2D and 3D problems by direct methods, such as the standard QR method, is complicated due to the large size of matrix $\mathcal{A}_k$, $n=4N_x N_y$, where $N_x$ and $N_y$ represent the number of grid points and the number of components in the fluctuation vector respectively.

Matrix-free methods, often known as time-stepper approaches, offer a solution for the leading eigenvectors and eigenvalues without explicitly constructing or solving the associated eigenmatrix. 
These methods involve solving the eigenvalue problem using snapshots of velocity fields at various time intervals, without storing matrices. 
Utilizing these time snapshots of velocity vectors, a lower-order eigenmatrix is created and solved, providing leading eigenvectors and eigenvalues for the original eigenmatrix.
This technique has gained popularity in both stability analysis \citep{barkley2002three, blackburn2008convective, bagheri2009global} and control design \citep{bagheri2009input}. 
Methods such as Arnoldi Iteration \citep{arnoldi1951principle, lehoucq1998arpack}, employed in the present work, facilitate this process. More details about the method can be found in 
\autoref{sec:matrixfreemethods}.

Solving \eqref{eq:simple_eigen2} for each spanwise wavenumber results in eigenfunctions with free amplitudes and phases, owing to its linear formulation. 
Prior to employing these eigenfunctions as acoustic source fluctuations through the ansatz provided in \eqref{eq:normal_zt}, it is necessary to calibrate their amplitudes. 
Once calibrated, these global mode fluctuations can be utilized as acoustic sources in an acoustic analogy to calculate far-field noise associated with them.

\subsection{Curle's acoustic analogy for a flow over a cylinder} \label{ssec:method_curle}

\subsubsection{Formulation} 

To compute sound radiation from global modes, an acoustic analogy can be used.
Lighthill's analogy \cite{lighthill1952sound, lighthill1954sound} reformulates the compressible Navier-Stokes equations into an inhomogeneous wave equation form,
\begin{equation}\label{eq:lighthill}
\left(  \frac{\partial^2}{\partial t^2} - c_0^2 \nabla^2 \right) \rho=\frac{\partial^2 \mathcal{T}_{ij}}{\partial x_i \partial x_j} (\mathbf{x},t).
\end{equation}
Here $\rho$ is the density, 
$c_0$ is the free-stream sound speed, 
$\mathcal{T}_{ij}=\rho u_i u_j+(p-c_0^2\rho)\delta_{ij} - \tau_{ij}$, incorporating velocity $u_i$, pressure $p$, and viscous stress components $\tau_{ij}$.
In \eqref{eq:lighthill}, the hydrodynamic flow-field data serves as the acoustic source on the right-hand side. To address the sound propagation problem, the inhomogeneous wave equation on the left-hand side is solved independently and separately from the hydrodynamics calculations or measurements.

Curle's analogy \citep{curle1955influence} includes the effects of bodies and surfaces in the flow and has been successfully applied to cylinder flows using a surface-dipole formulation (for instance, see \cite{inoue2002sound, gloerfelt2005flow, margnat2015hybrid}).
Derived from Lighthill's analogy, Curle's analogy for a steady impermeable surface allows the radiated sound pressure to be expressed as,
\begin{equation}\label{eq:curles_final}
\rho'(\mathbf{x},t)=\frac{\partial^2 }{\partial x_i \partial x_j} \iiint_{V\llap{--}} \left[ \frac{\mathcal{T}_{ij}}{4\pi c^2 \mid \mathbf{x}-\mathbf{w} \mid } \right] d^3\mathbf{w} - 
\frac{\partial }{\partial x_i} \iint_{\partial V\llap{--}}
\left[ \frac{p'n_i}{4\pi c^2 \mid \mathbf{x}-\mathbf{w} \mid } \right] d^2\mathbf{w},
\end{equation}
where, $\rho'=\rho-\rho_0, p'=p-p_0$ with 
$(\rho_0, p_0)$ representing the constant reference state in the observer domain, 
$\partial {V\llap{--}}$ 
represents the boundary of the control volume $V\llap{--}$ and the viscous stress has been neglected in the surface term. The bracketed terms are to be evaluated at the retarded (emission) time, $\tau^* = t-\mid \mathbf{x} - \mathbf{w} \mid/c$. 
The first part of the R.H.S. in \eqref{eq:curles_final} corresponds to the incident field, a quadrupole source in the free field,
while the second part corresponds to the scattered field~\citep{gloerfelt2005flow}, a dipole source in the free field.
In the frequency domain, \autoref{eq:curles_final} reads
\begin{equation}\label{eq:intergral_solution_freq}
\hat{p}(\mathbf{x},\omega)=
c^2 \hat{\rho}(\mathbf{x},\omega)=
-\iiint_{V\llap{--}} \frac{\partial^2 \hat{G}}{\partial w_i \partial w_j} \hat{T}_{ij} \hspace{0.5mm} d^3 \mathbf{w} \hspace{0.5mm}
-\iint_{\partial V\llap{--}} \frac{\partial \hat{G}}{\partial w_i} \hat{p}(\mathbf{w},\omega) n_j \hspace{0.5mm} d^2 \mathbf{w} \hspace{0.5mm},
\end{equation}
where $\hat{G}(\mathbf{x} \mid \mathbf{w}, \omega)$ is the free-field Green function in the frequency domain, given by
\begin{equation}
\hat{G}(\mathbf{x} \mid \mathbf{w}, \omega) = -\frac{e^{-i k_0 \mid \mathbf{x} - \mathbf{w} \mid}}{4\pi \mid \mathbf{x} - \mathbf{w} \mid}
\end{equation}
where $k_0=\omega/c_0 = 2 \pi St U_{\infty} /c_0d$ is the acoustic wavenumber.

\subsubsection{Acoustic compactness of the cross section} 
\label{sssec:acoustic_compactness}
Acoustic compactness connects the sound emission characteristic length i.e. the acoustic wavelength ($\lambda$) with the cylinder's characteristic length.
If the diameter of the cylinder is negligible in comparison to an acoustic wavelength, $\lambda \gg d $, the differences in retarded time for the different locations of the cylinder across its circumference are negligible and the cylinder may be considered as a point source in the sectional plane.
If the spanwise characteristic length is much smaller than the acoustic wavelength ($\lambda \gg L$ or $L_c$), then the differences in the retarded time for various points along the span of the cylinder are negligible. Otherwise, these differences need to be considered.

For the application of noise model for the Re=13300 flow over a cylinder, which will be presented in Section \ref{sec:results_re13300}, we have  $d = 0.01$ m, $L = 0.7$ m and $U_{\infty} = 20$ m/s which corresponds to Mach number, $M = U_{\infty}/c_0 = 0.06$. The spanwise coherence length for this flow has been measured as $L_c \simeq 5d=0.05$m \citep{margnat2023cylinder}. 
The Aeolian tone frequency, as measured during the present experiment campaign and which will be presented in Section \ref{sssec:micro_setup}, is $f \approx 400$ Hz which corresponds to an acoustic wavelength $\lambda \approx 0.85$ m.
Across section, compactness ratio $r_{\text{com}}^{\text{sec}}=\lambda/d=85$, meaning that the cylinder can be assumed compact across its section. Across span, if the cylinder length is taken as spanwise characteristic length, compactness ratio , $r_{\text{com}}^{\text{span}}=\lambda/L \approx 1.2$, meaning that the span of the cylinder is acoustically compact. However, if the coherence length is taken as spanwise characteristic length, then  $r_{\text{com}}^{\text{span}}=\lambda/L_c \approx 17$. 

If the cylinder is taken as an overall compact source, an order of magnitude analysis for \autoref{eq:intergral_solution_freq} shows that the quadrupolar contribution of the volume integral to the sound power is $M^2$ smaller than the contribution of the dipolar surface integral \citep{dowling1984sound}.
Thus for $M \ll 1$ flows, the dipole noise contribution shall be much stronger than the quadrupolar noise contribution.
Hence, for such flows, the quadrupole source can be neglected, giving
\begin{equation}
\hat{p}(\mathbf{x},\omega)= -\iint_{\partial V\llap{--}} \frac{\partial \hat{G}}{\partial w_i} \hat{p}(\mathbf{w},\omega) n_j \hspace{0.5mm} d^2 \mathbf{w} \hspace{0.5mm},
\end{equation}
where
\begin{equation}\label{eq:dgdy}
\frac{\partial G}{\partial w_i}=
 \frac{e^{-i k_0 r_i}}{4\pi r} \left(  \frac{r_i}{r^2} +\frac{ik_0 r_i}{r} \right),
\end{equation}
where $r_i=x_i-w_i$, $r=\mid \mathbf{x} - \mathbf{w} \mid$. This gives,
\begin{equation} \label{eq:curle_10}
\hat{p}(\mathbf{x},\omega)= -\int_{-L/2}^{L/2} \oint_{C} \hat{P}(w_1,w_2,w_3,\omega)  \hspace{0.5mm}
\frac{e^{-i k_0 r}}{r}  \hspace{0.5mm}
\frac{r_i}{4\pi r}  \left(  {\frac{1}{r} + ik_0} \right) \hspace{0.5mm}
n_i \hspace{0.5mm}
\mathrm{d}l \hspace{0.5mm} \mathrm{d}w_3,
\end{equation}
where $\hat{P}(w_1,w_2,w_3,\omega)$ is the hydrodynamic pressure fluctuation distribution over the cylinder surface $(w_1,w_2,w_3)$ for the considered frequency, $\omega$,
$C$ is the closed loop on cylinder surface for the $(w_1,w_2)$ plane with
$\mathrm{d}l$ the elementary length along $C$.


For the present application of the Re=13300 flow over a cylinder, as which will be presented in Section \ref{sssec:gmc_re13300}, only spanwise-homogeneous type fluctuations are considered which gives $\hat{p}(w_1,w_2,w_3,\omega)=\hat{p}(w_1,w_2,\omega)$.
\autoref{eq:curle_10} is further simplified by using the sectional compactness assumption $(\sqrt{x_1^2+x_2^2} \gg \sqrt{w_1^2+w_2^2})$, that allows to use $r_1=x_1-w_1\approx x_1$, $r_2=x_2-w_2\approx x_2$ and $r_3=x_3-w_3$. Taking these two factors into account, we finally arrive at 
\begin{equation} \label{eq:curle_11}
\hat{p}(\mathbf{x},\omega)= 
-\int_{-L/2}^{L/2} \frac{e^{-i k_0 r}}{r}  \hspace{0.5mm}
\frac{\left(x_3-w_3\right)}{4\pi r}  \left(  {\frac{1}{r} + ik_0} \right) \mathrm{d}w_3 \times
\oint_{C} \hat{P}(w_1,w_2,\omega)  \hspace{0.5mm}
n_i \hspace{0.5mm}
\mathrm{d}l \hspace{0.5mm} ,
\end{equation}
where $r=\sqrt{x_1^2+x_2^2+(x_3-w_3)^2}$. The first integral on the R.H.S. of \eqref{eq:curle_11} is across the cylinder length, hence not assuming spanwise compactness here i.e. retaining the retarded time effect across the span. The second integral is the resulting, sectional pressure force, which can be evaluated independently of the acoustic estimation.

Summarising the noise modelling approach, the global modes, ($\widetilde{\mathbf{q}}, k,\omega$), are obtained from linear global stability analysis on the mean flow. After the amplitude calibration of the global modes, they are used, via ansatz \eqref{eq:normal_zt}, to get the pressure fluctuations distribution over the cylinder surface, $\hat{P}(w_1,w_2,w_3,\omega)$.
These surface pressure fluctuations are then utilized in \eqref{eq:curle_11}, for the calculation of sound radiation associated with these global modes.
Section \ref{sec:results_re150} and Section \ref{sec:results_re13300} present the implementation of the proposed noise model for flows at Re=150 and Re=13300, respectively, over a cylinder.

\section{Validation: Re=150 flow over a cylinder}\label{sec:results_re150}

In this section, the noise  model is built for a case of Re=150 flow over a circular cylinder, for validation purposes. 
Opting for this low-Re flow was driven by existing literature and the advantage of more regular and easier-to-analyze cylinder wakes compared to the turbulent wakes at high-Re.

\subsection{Mean flow evaluation}\label{ssec:re150_mean}

An incompressible 2D DNS is done for the present configuration to record the time-resolved flow field which was then used to evaluate the time-averaged velocity and pressure fields. 
It employs the Nektar++ solver \citep{cantwell2015nektar++} which is based on the spectral/hp element method which combines the geometric flexibility of classical h-type finite element techniques with the desirable resolution properties of spectral methods by increasing the polynomial order (p-type) in regions demanding higher accuracy.
These techniques have been applied in many fundamental studies of fluid mechanics \cite{sherwin1996tetrahedralhpfinite}.

Flow setup is 2D and is same as shown in the Figure \ref{fig:sketch}(a).
The computational domain used for the numerical simulation is kept as $-45 <x/d< 125, -45 <y/d <45$.

\begin{figure}[b]
     \centering
\begin{subfigure}[b]{0.67\textwidth}
         \centering
	\includegraphics[height=4cm]{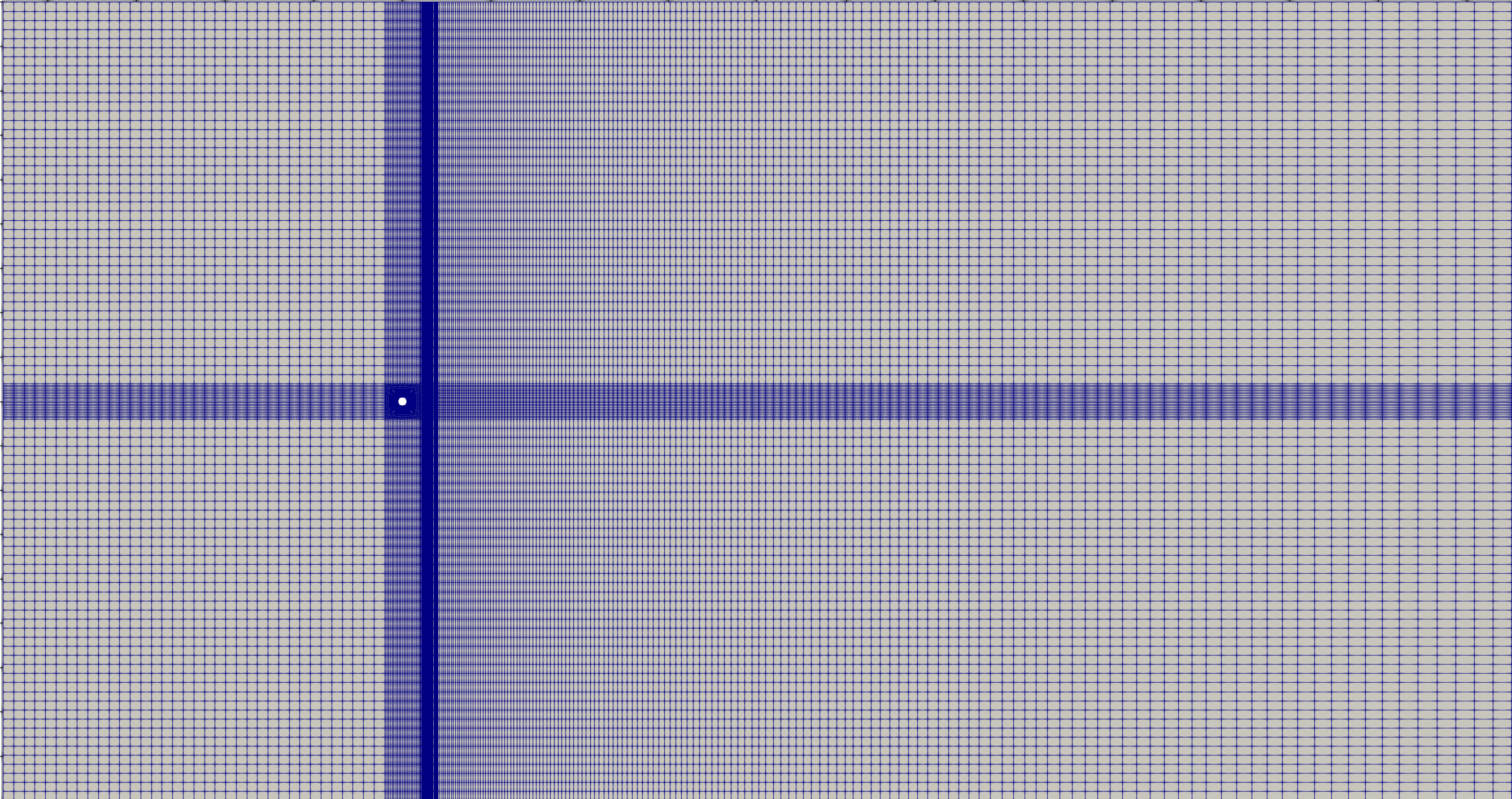}
        \caption{Global view. Domain: $-45<x/d<125$, $-45<y/d<45$.}
\end{subfigure}
     \hfill
\begin{subfigure}[b]{0.3\textwidth}
	\includegraphics[height=4cm]{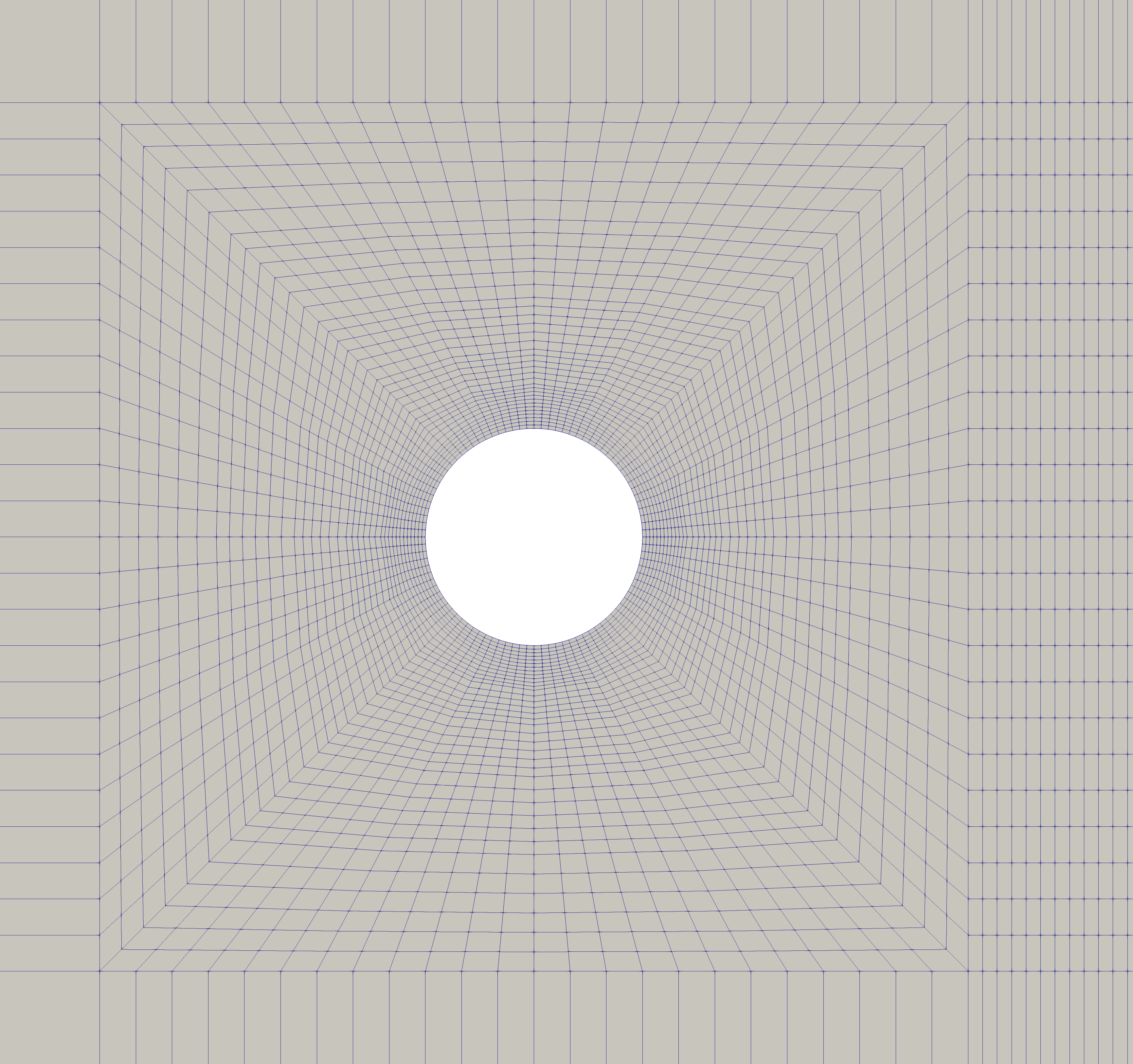}
        \caption{Closer view.}
\end{subfigure}
        \caption{Re=150 flow over a circular cylinder: Computational grid.}
        \label{fig:re150 domain}
\end{figure}

An unstructured grid is used with mesh refinement near the cylinder ($\Delta x/d = 0.017$) and a coarser mesh is used at the domain boundaries ($\Delta x/d = 1.2$), which can be seen in \autoref{fig:re150 domain}(a). 
The polynomial expansions used on grid points were of the order ``5".
No-slip conditions are imposed on the cylinder.
Uniform flow, $(U, V) = (U_{\infty}, 0)$, is imposed upstream and an outflow boundary condition is imposed at the downstream end of the domain and on the lateral sides. The time-step, $\Delta t\times U_{\infty}/d=0.001$ was used for the time-evolution of the solution. This time-step was chosen according to the Courant–Friedrichs–Lewy (CFL) condition and close to the time-step ($\Delta t\times U_{\infty}/d=0.002$) in 2D compressible DNS for Re=150 flow by \citet{inoue2002sound}.

\begin{figure}
     \centering
\begin{subfigure}[b]{0.48\textwidth}
         \centering
	\includegraphics[width=1\textwidth]{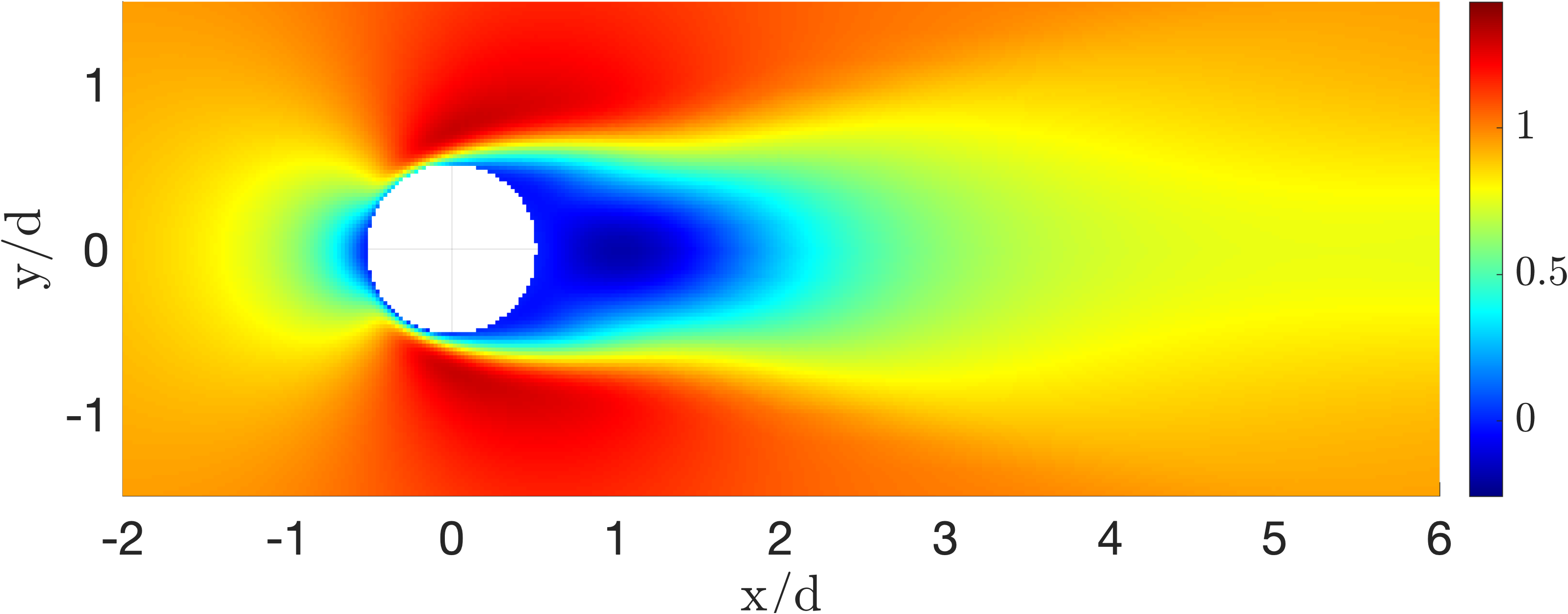}
        \caption{$\overline{U}/U_{\infty}$, mean streamwise velocity}
\end{subfigure}
     \hfill
\begin{subfigure}[b]{0.48\textwidth}
	\includegraphics[width=1\textwidth]{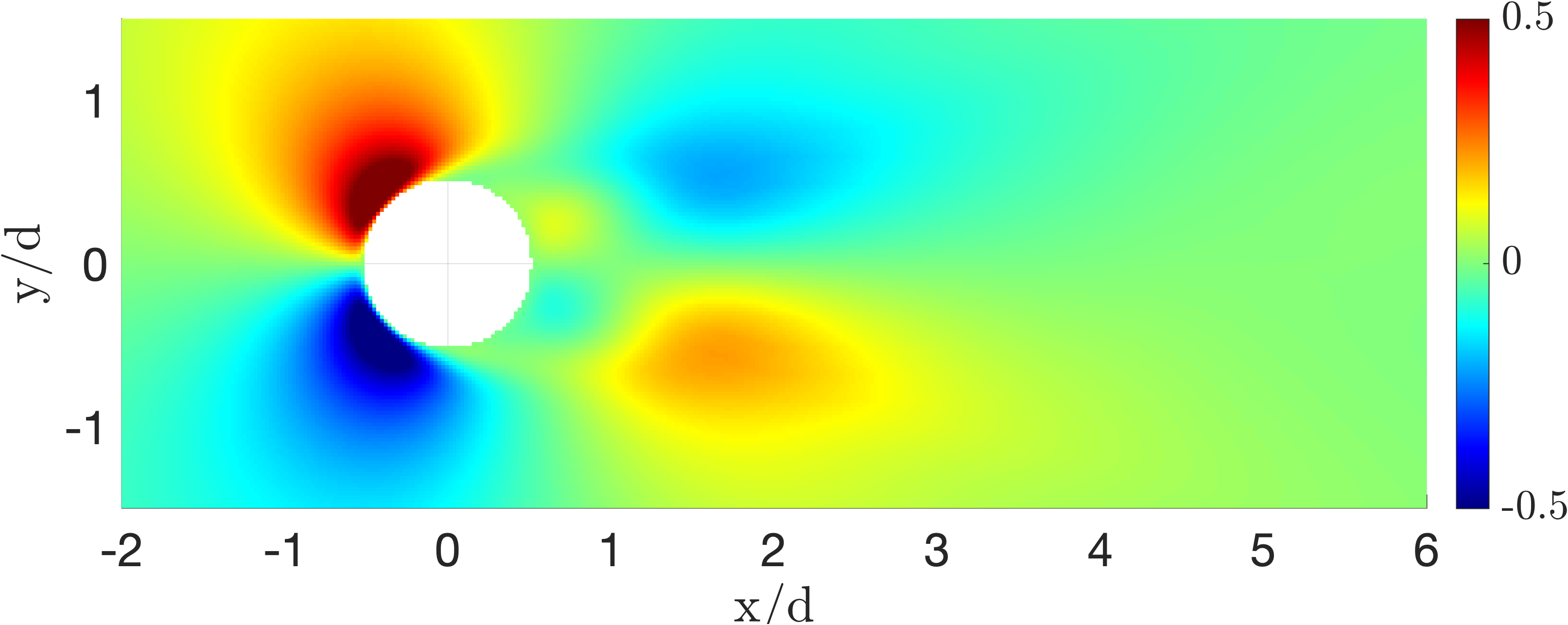}
        \caption{$\overline{V}/U_{\infty}$, mean transverse velocity}
\end{subfigure}
	\vfill
\begin{subfigure}[b]{0.48\textwidth}
	\includegraphics[width=1\textwidth]{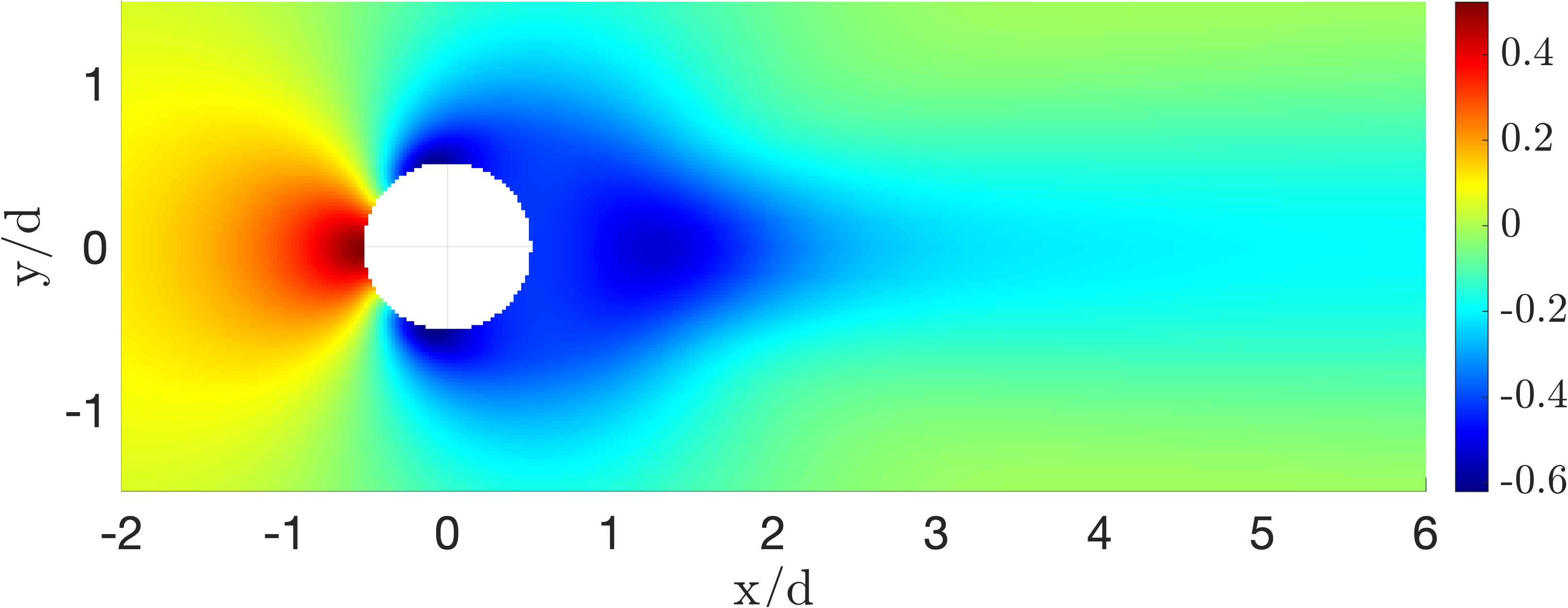}
        \caption{$\overline{P}/\rho U_{\infty}^2$, mean pressure}
\end{subfigure}
        \caption{Re=150 flow over a circular cylinder: mean flow from DNS.}
        \label{fig:re150 mean}
\end{figure}

After reaching the developed flow state, the flow is evolved for 20 vortex-shedding cycles that signify the total time, $T\times U_{\infty}/d=104$ (Total number of time-steps, $N_t=130 000$) to evaluate the time-averaged flow field, which is presented in \autoref{fig:re150 mean}. 

\subsection{Global stability analysis} \label{ssec:gsa_re150}
The linear global stability analysis for the time-averaged mean flow for the Re=150 flow around a circular cylinder is done here to get the leading global modes, considering only $k=0$ spanwise mode.

The same grid as during the mean flow calculation but with the polynomial expansions on grid points being of the order ``7", is used here, which corresponds to 38318 grid points distributed over the domain of $-45<x/d<125, -45<y/d<45$.
Boundary conditions are kept as zero velocity fluctuations at the cylinder surface, left boundary, top boundary and bottom boundary, and outflow at the right boundary. The Krylov space of $\kappa=256$ is used for global stability analysis.

\begin{figure}
     \centering
	\includegraphics[width=0.6\textwidth]{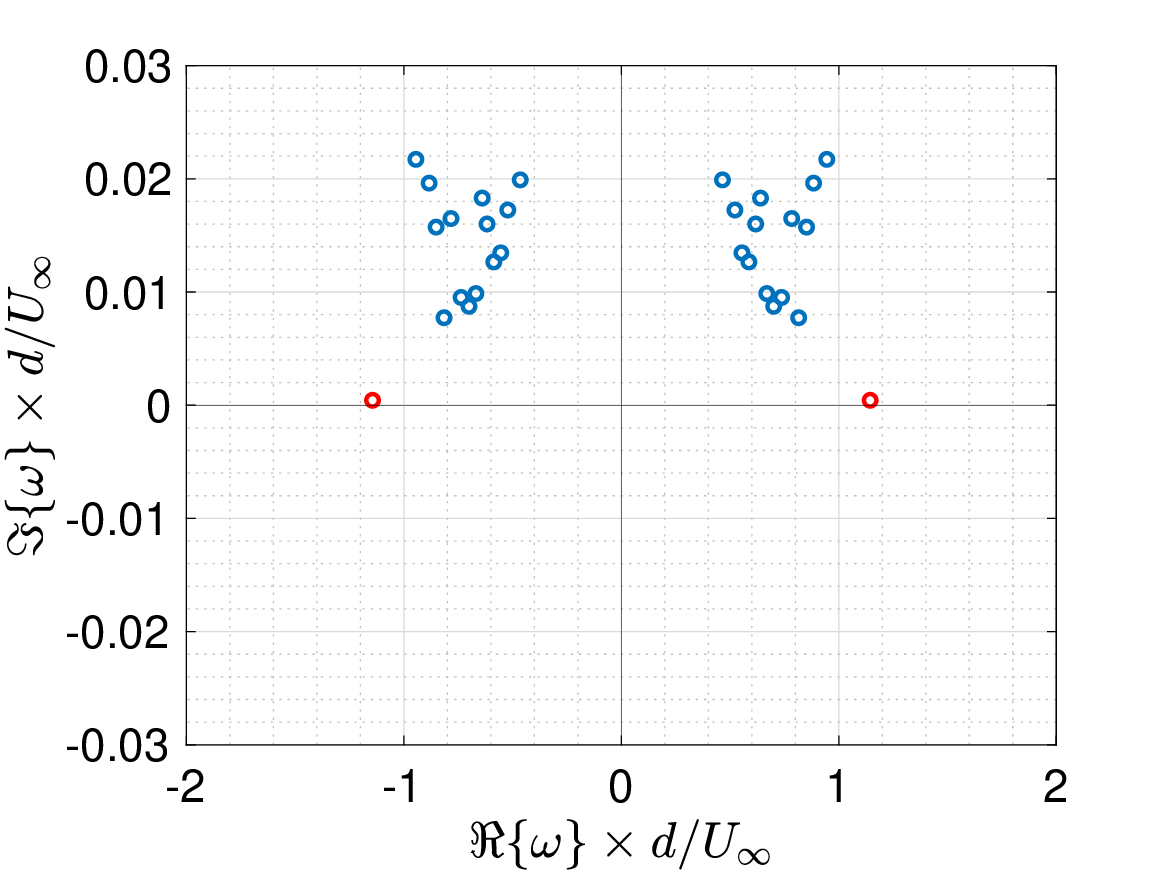}
        \caption{Re=150 flow over a circular cylinder: Eigenspectrum for the leading 30 global stability modes. All modes are stable with 2 modes being neutrally stable,  marked in red.}
        \label{fig:re150 eigenspectrum}
\end{figure}

Among the converged leading eigenmodes (largest growth rates), eigenvalues for the leading 30 global modes are presented in the eigenspectrum as shown in \autoref{fig:re150 eigenspectrum}.
Here, the eigenfunctions corresponding to a mode in the left plane of the eigenspectrum, together with its counterpart in the right plane, represent the real and imaginary parts of the global mode eigenfunction, forming a pair.

Only one pair of modes in \autoref{fig:re150 eigenspectrum} (marked in red circles) is found to be nearly neutrally stable, all other pairs were stable.
The eigenvalues corresponding to this leading global mode are 
\begin{equation}
\omega \times d/U_{\infty}= \pm 1.1443 + 0.0004 i,
\end{equation}
indicating that the mean flow is marginally stable, consistent with previous studies \citep{mantivc2014self, barkley2006linear, fani2018computation}.
The fluctuation frequency for this mode corresponds to the Strouhal number St$=fd/U_{\infty}=\Re \{ \omega \} /2\pi \times d/U_{\infty}=0.182$. This value is in agreement with St = 0.183 for the lift fluctuation (LF) frequency in the compressible DNS by \citet{inoue2002sound}.

\begin{figure}
     \centering
\begin{subfigure}[b]{0.48\textwidth}
         \centering
	\includegraphics[width=1\textwidth]{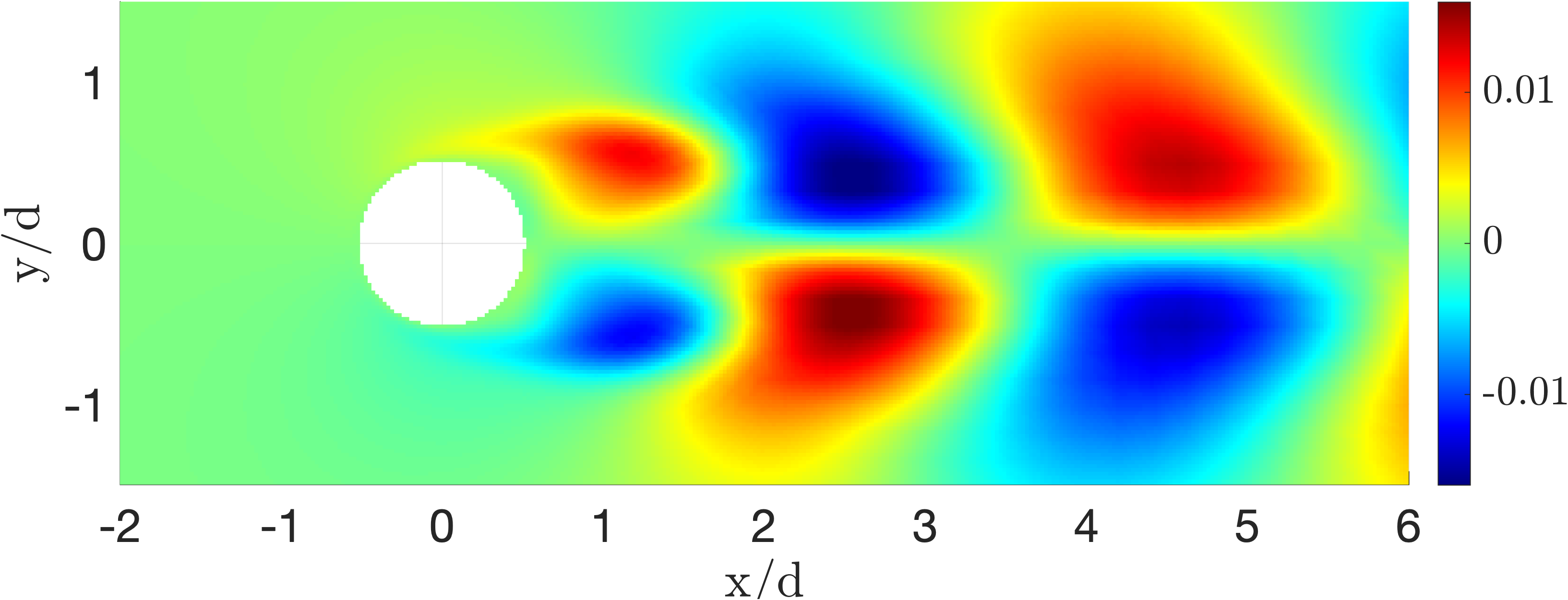}
        \caption{$	\Re\{ \tilde{u} \}/U_{\infty}$, streamwise velocity fluctuations}
\end{subfigure}
     \hfill
\begin{subfigure}[b]{0.48\textwidth}
         \centering
	\includegraphics[width=1\textwidth]{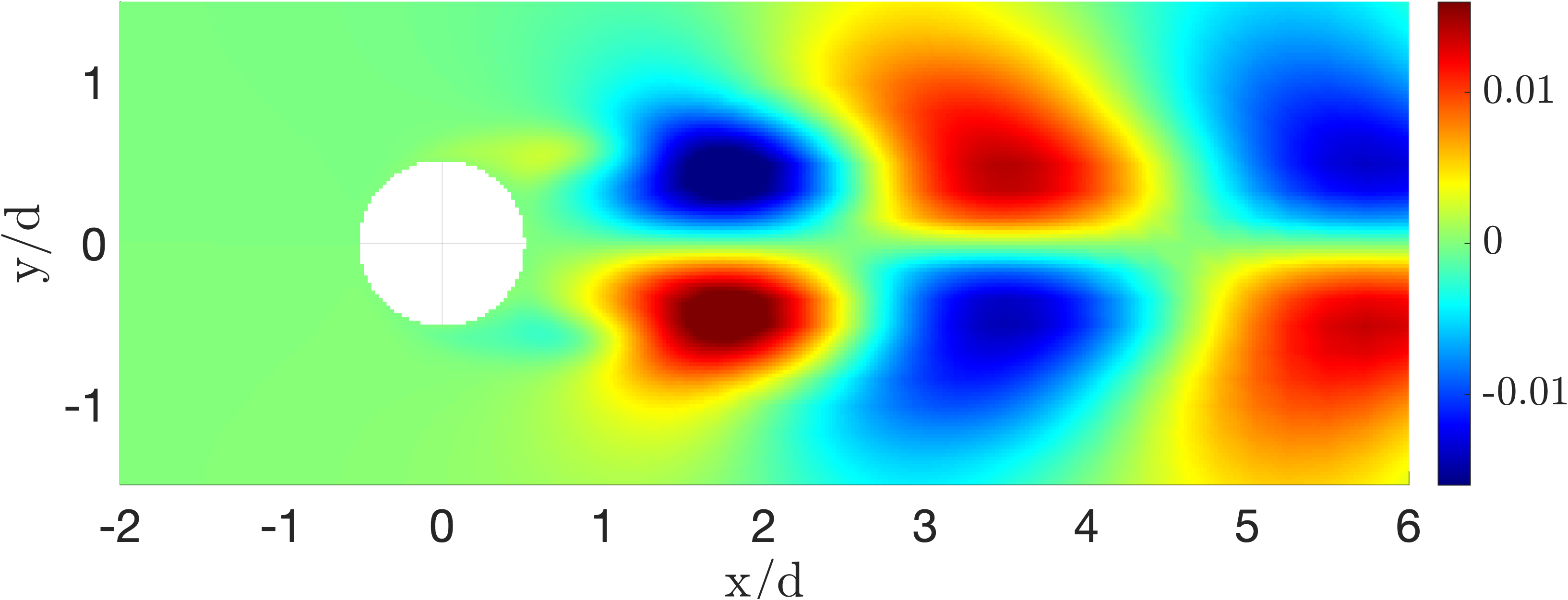}
        \caption{$	\Im\{ \tilde{u} \}/U_{\infty}$,  streamwise velocity fluctuations}
\end{subfigure}
     \vfill
\begin{subfigure}[b]{0.48\textwidth}
	\includegraphics[width=1\textwidth]{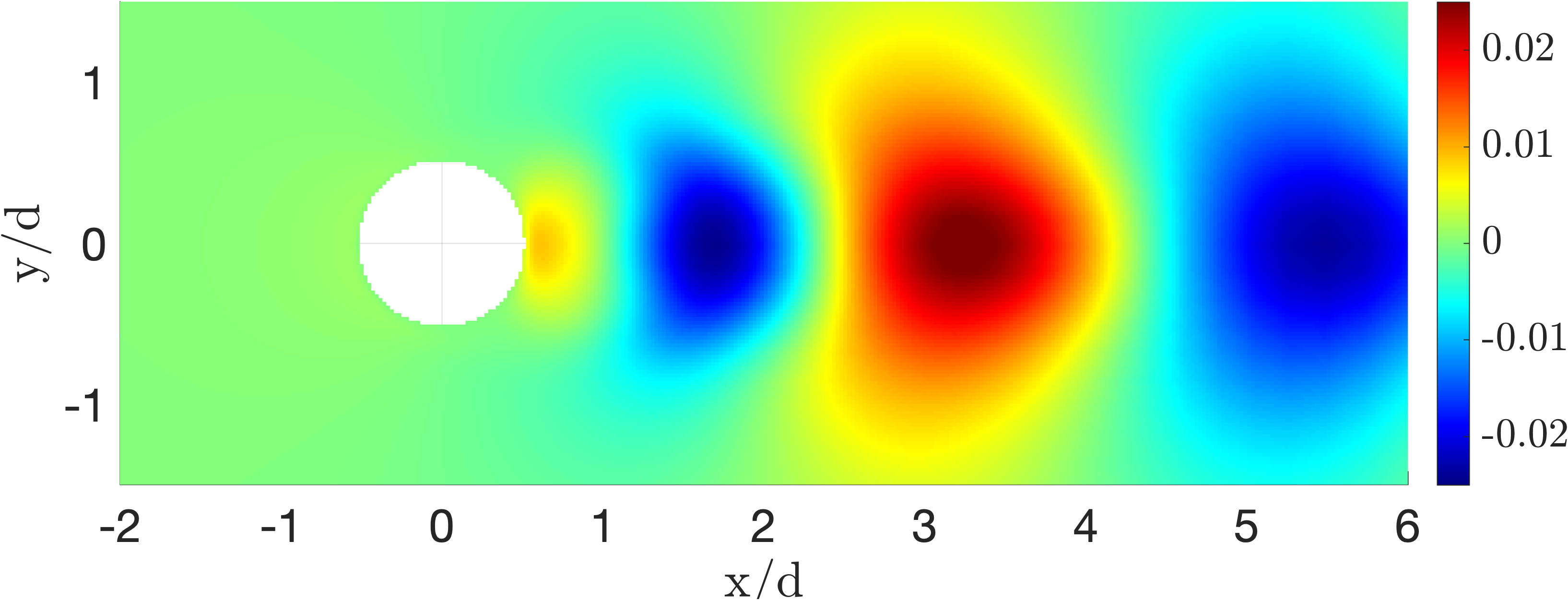}
        \caption{$	\Re\{ \tilde{v} \}/U_{\infty}$,  transverse velocity fluctuations}
\end{subfigure}
	\hfill
\begin{subfigure}[b]{0.48\textwidth}
	\includegraphics[width=1\textwidth]{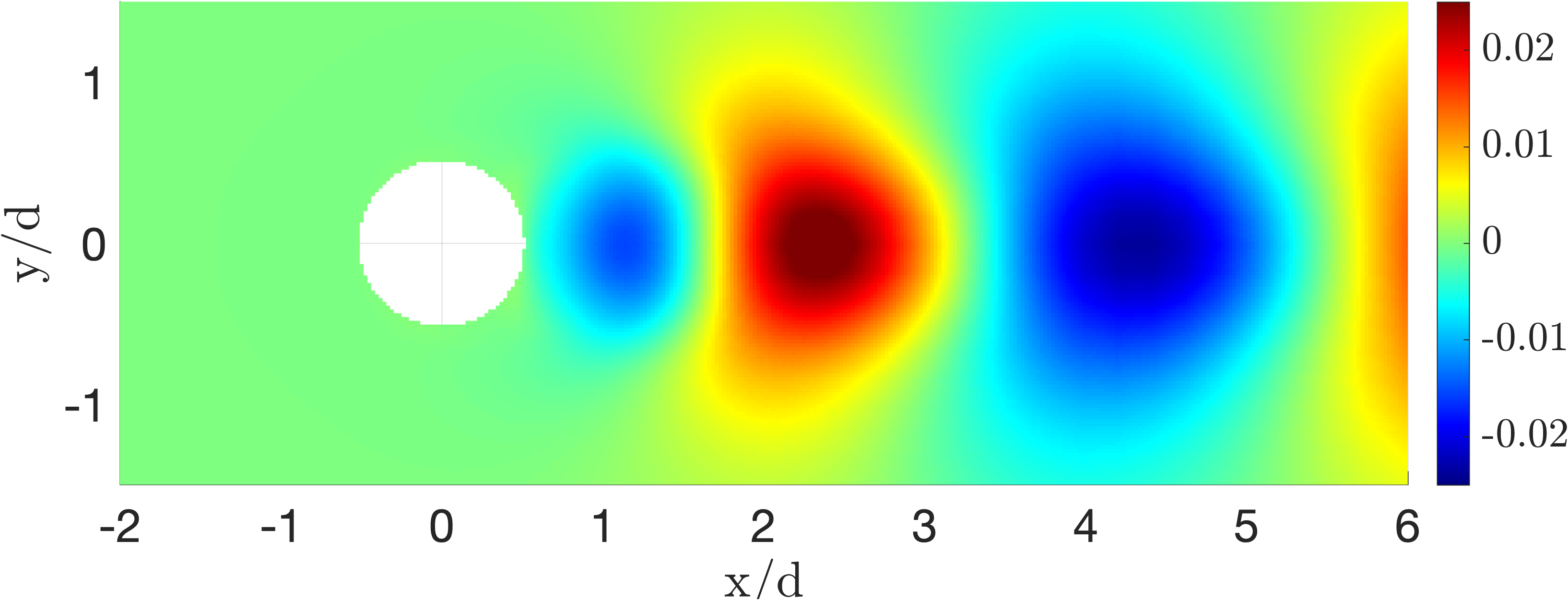}
        \caption{$	\Im\{ \tilde{v} \}/U_{\infty}$,  transverse velocity fluctuations}
\end{subfigure}
	\vfill
\begin{subfigure}[b]{0.48\textwidth}
	\includegraphics[width=1\textwidth]{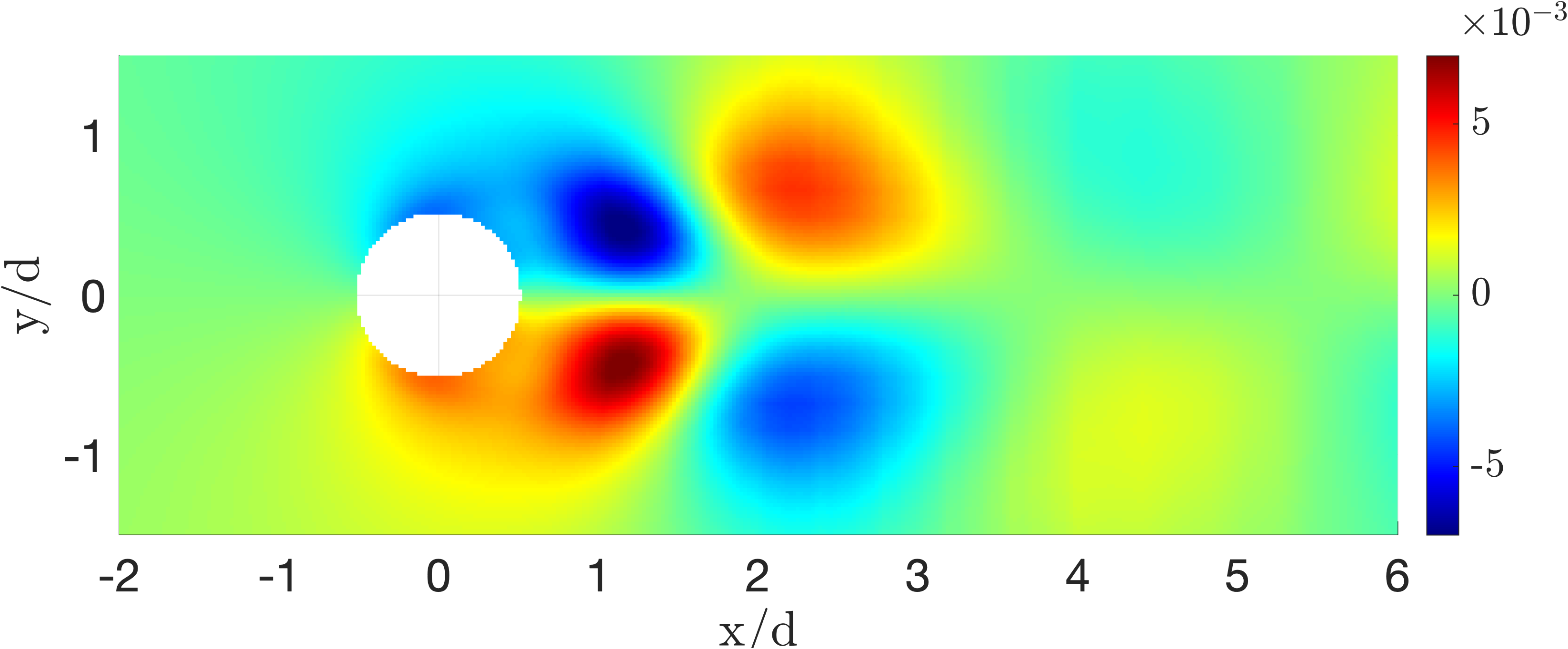}
        \caption{$	\Re\{ \tilde{p} \}/\rho U_{\infty}^2$,  pressure fluctuations}
\end{subfigure}
	\hfill
\begin{subfigure}[b]{0.48\textwidth}
	\includegraphics[width=1\textwidth]{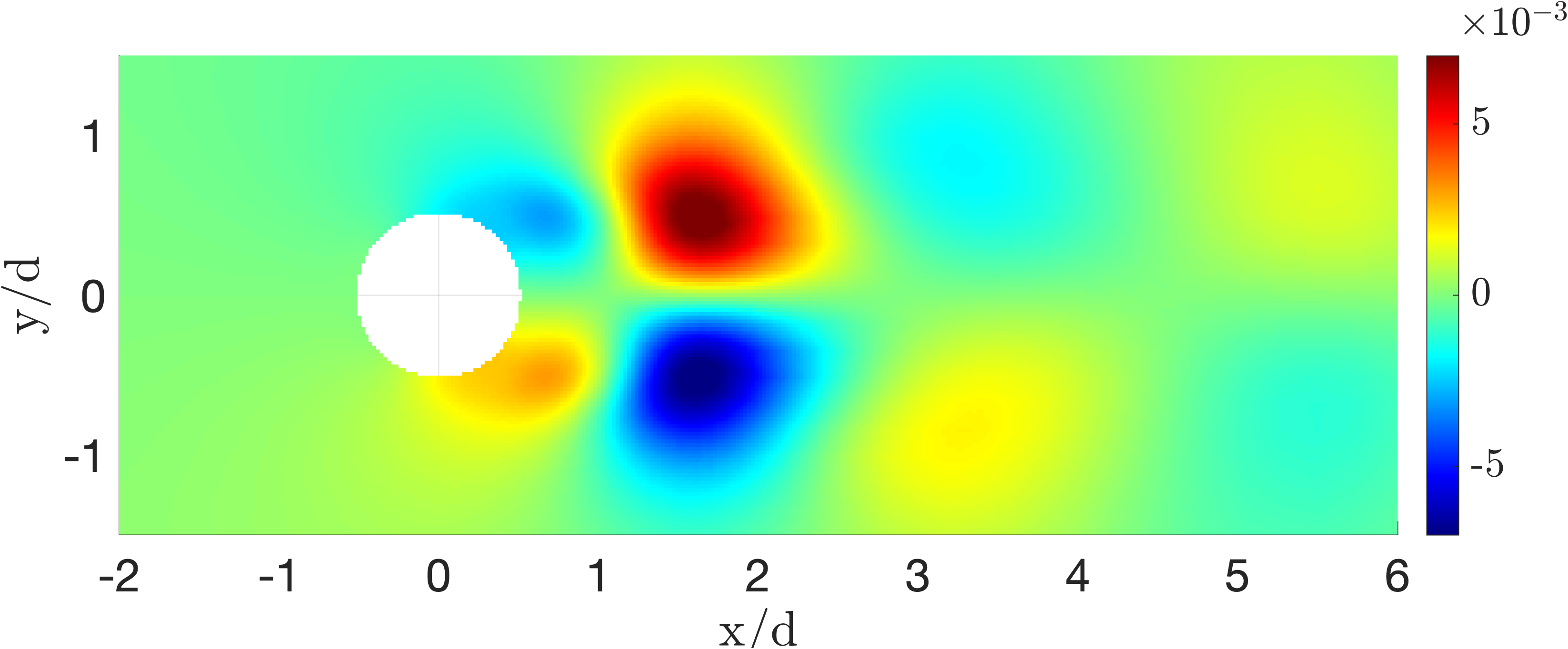}
        \caption{$	\Im\{ \tilde{p} \}/\rho U_{\infty}^2$,  pressure fluctuations}
\end{subfigure}
        \caption{Re=150 flow over a circular cylinder: structure of the leading eigenmode ($\Re \{ \omega \} /2\pi \times d/U_{\infty}=0.182$).}
        \label{fig:re150 eigenstructures}
\end{figure}

The spatial structure of the mode can be seen in the \autoref{fig:re150 eigenstructures}. 
It shows alternate vortices in $\pm y$ positions, being propagated downstream with the mean flow while growing at the beginning as they travel downstream and reaching a saturated state. Not shown here in \autoref{fig:re150 eigenstructures}, but beyond $x \approx 30d$ downstream of the body, they tend to decay.
The eigenstructures are also found to be in qualitative agreement with the eigenstructures from the global stability done by \citet{fani2018computation} at this Re. 

Also, the domain size dependence on the leading global mode was checked by performing global stability analysis on various domains and comparing their eigenvalues and eigenvectors together. 
The quantitative similarity of the eigenvectors from domains with that of domain A is calculated by
the function, $\mid \tilde{p} \tilde{p}_A^H \mid / \mid \tilde{p}\mid  \mid \tilde{p}_A \mid $, where $ \tilde{p}$ and $\tilde{p}_A $ are the complex eigenfunctions on the cylinder surface for that particular domain and domain A respectively, and $H$ represents the Hermitian transpose.
The results are shown in \autoref{tab:eigenre150_domain_convergence}. 
We see that the fluctuation frequency for the dominant global mode and eigenfunctions are very well converged.
The growth rate ($\Im \{ \omega \}$) for the mode is also close to zero for all the domains considered.

\begin{table*}
        \centering
                \caption{Domain convergence test for dominant mode from global stability analysis for Re=150 flow over a circular cylinder. Eigenvalues and eigenfunctions at the cylinder surface are compared.}
        \centering
\label{tab:eigenre150_domain_convergence}
\begin{tabular}{@{}lllll@{}}
\toprule
Domain & $x/d$  & $y/d$ & $\omega$ & ${\mid \tilde{p} \tilde{p}_A^H \mid }/{ \mid \tilde{p}\mid  \mid \tilde{p}_A \mid }$ \\
\midrule
A  \hspace{2mm} & -45 to +125 & -45 to +45& $\pm 1.1443  + 0.0004 i$& 1 \\
B  \hspace{2mm} & -16 to +46 & -20 to +20&  $\pm 1.1446  + 0.0009 i$& 0.9964\\
C  \hspace{2mm} & -9 to +32 & -14 to +14& $\pm 1.1433  + 0.0004 i$ & 0.9962\\
D  \hspace{2mm} & -2 to +19 & -8 to +8& $ \pm 1.1540  - 0.01 i$& 0.9583\\
E  \hspace{2mm} & -2 to +10 & -8 to +8 & $ \pm 1.1540  - 0.01 i$& 0.9723\\
F  \hspace{2mm} & -2 to +10 & -2 to +2& $\pm 1.1707  - 0.03 i$ & 0.9833\\
G  \hspace{2mm} & -2 to +8 & -2 to +2& $\pm 1.1708  - 0.03 i $& 0.9575\\
H  \hspace{2mm} & -2 to +6 & -2 to +2& $ \pm 1.1710  - 0.03 i$& 0.9582\\
I  \hspace{2mm} & -2 to +4 &-2 to +2 &$ \pm 1.1697  - 0.03 i $& 0.9616\\
\botrule
\end{tabular}
\end{table*}

\subsection{Mode calibration}\label{ssec:re150_calib}
\begin{figure}
     \centering
	\includegraphics[width=0.6\textwidth]{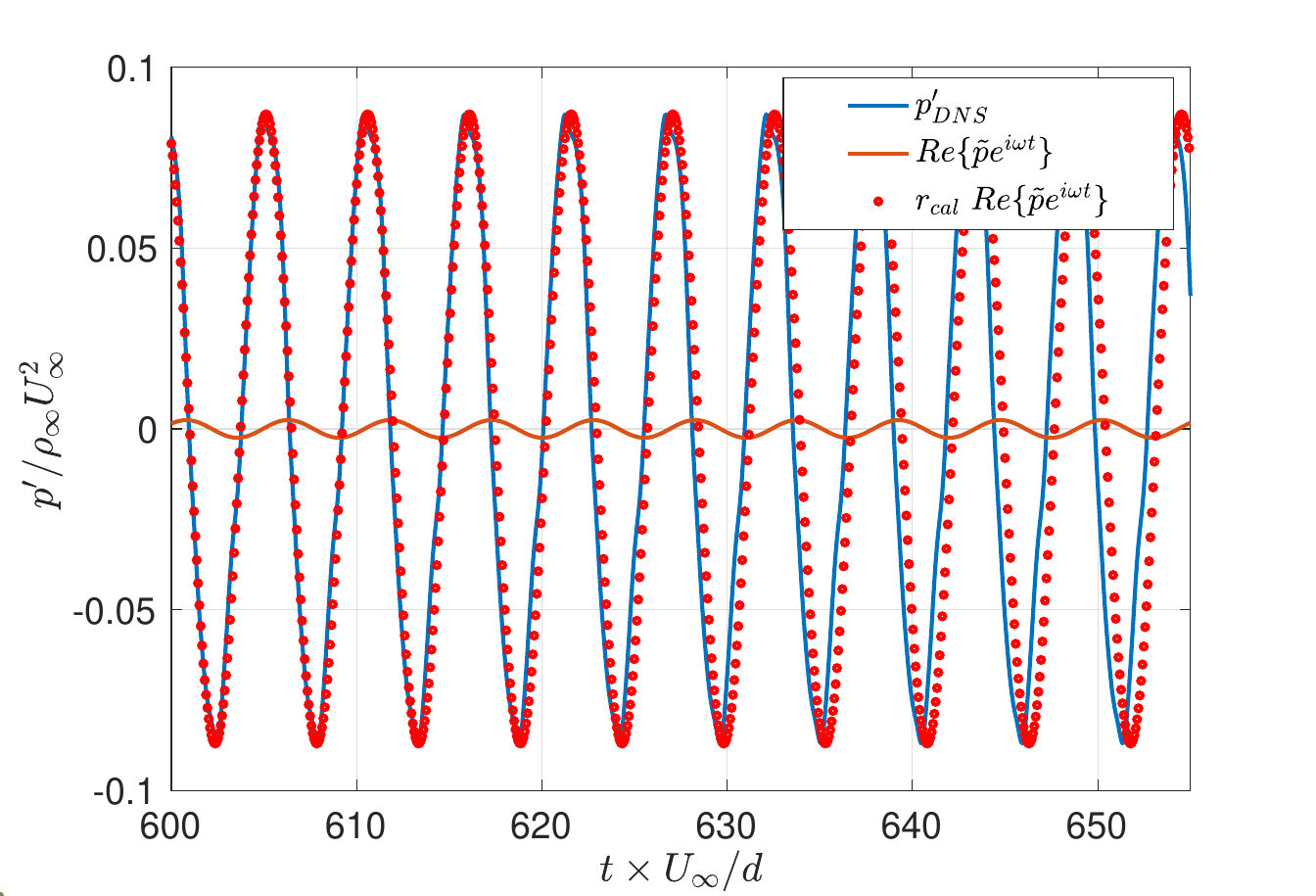}
        \caption{Re=150 flow over a circular cylinder: Global mode calibration.}
        \label{fig:re150 calibration}
\end{figure}

As the linear global modes have a free amplitude, this is calibrated here using a single-point signal from the DNS. The time history of pressure fluctuations, at a single spatial location, is recorded during the incompressible DNS run, to calibrate the LF global mode.
For the probe location, $x/d=0, y/d=0.9$, the pressure time-series from DNS is shown in \autoref{fig:re150 calibration} for a time-snippet. 
The global-mode amplitude is adjusted to match the DNS, to give the comparison shown in \autoref{fig:re150 calibration}.
The robustness of the calibration method was checked by 
recording the pressure fluctuations, $p$, at various spatial locations in the region $-1<x/d<3,0.1<y/d<1.5$. 
In the considered region, the change in calibration ratio was found to be within $ \pm 22\%$.

\subsection{Farfield noise by the global mode} \label{ssec:re150_noisePredictions}
The far-field noise produced by the calibrated global mode is evaluated here by employing Curle's analogy \eqref{eq:curle_11}.
The noise directivity calculations are done along a circular arc of radius $r/d=75$ centred at the cylinder centre, 
where the 2D global mode pressure fluctuations were homogeneously distributed along the cylinder’s spanwise length of $10^4$ diameters. This length is selected as a numerical approximation to the infinitely long span of the cylinder, simulating a 2D case. 
This choice enables comparison of our noise calculations with the 2D compressible DNS conducted by \citet{inoue2002sound}.

\begin{figure}
     \centering
	\includegraphics[width=0.6\textwidth]{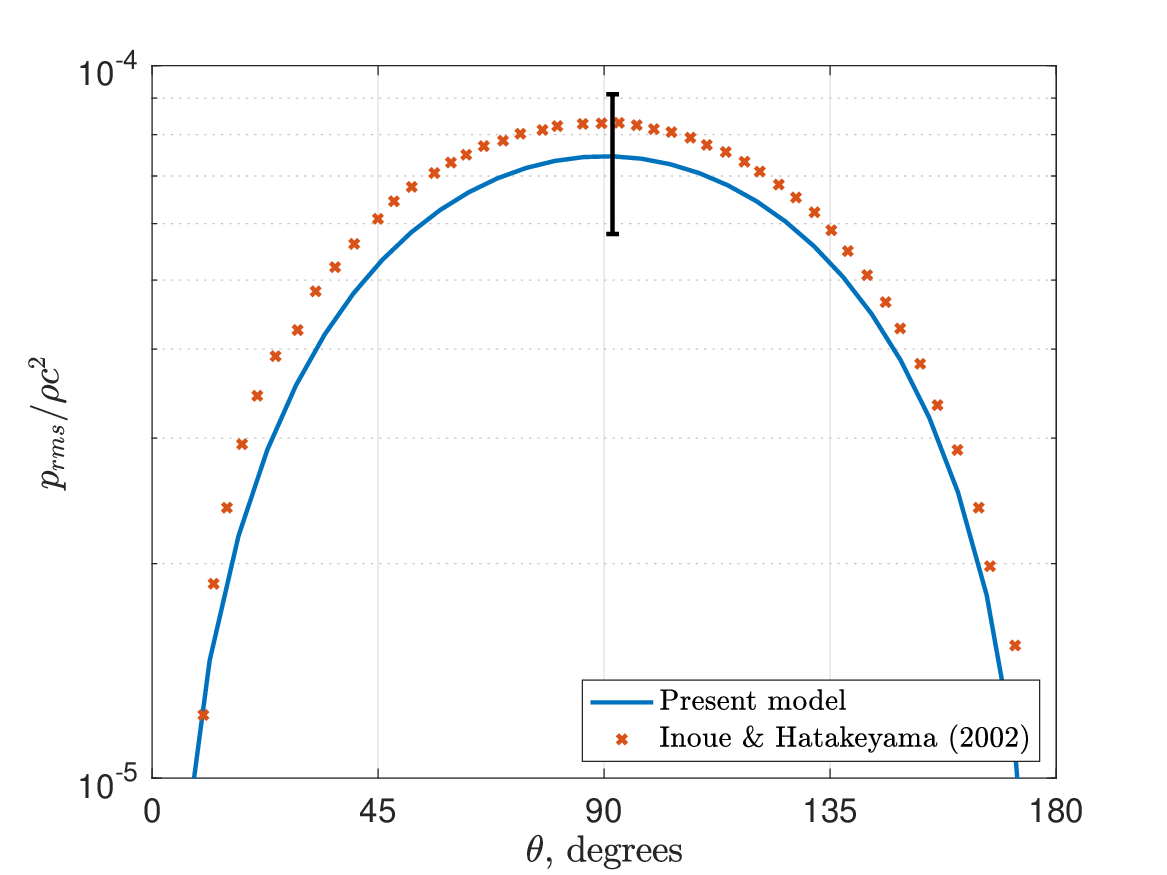}
        \caption{Re=150 flow over a circular cylinder: Far-field sound directivity at $r/d=75$. Streamwise direction
corresponds to $\theta=0$.
The error bar at $\theta=90$ (in black) corresponds to the $ \pm 22\%$ variation in calibration ratio.}
        \label{fig:re150 noise prediction}
\end{figure}

The sound directivity at $r/d=75$  is shown in \autoref{fig:re150 noise prediction}, exhibiting its dipole nature, as expected from Curle’s approach for such a low-drag geometry.
\autoref{fig:re150 noise prediction} also shows the comparison with the output of a compressible DNS for Re=150, Mach number, $M=0.2$ \cite{inoue2002sound}. 
Note that the $ \pm 22\%$ variation in calibration ratio leads to $\pm 2$ dB variations in the far-field sound. The median underestimates the DNS value by about $1$ dB, but these calibration bounds contain it.

\section{Application: Re=13300 flow over a cylinder}\label{sec:results_re13300}


The noise model is now applied to a Re=13300 flow over a spanwise-homogeneous circular cylinder in this section. 
To this aim, flow and acoustic measurements have been performed, specifically, (i) XY Mean PIV: time-averaged mean velocity measurements in the streamwise-transverse (XY) plane at the mid-span position ($z=0$), which will be used to perform stability analysis, (ii) XZ TR PIV: Time-resolved velocity measurements in a  streamwise-spanwise (XZ) plane, which will be used to extract the fluctuation amplitude of the spanwise Fourier modes of interest for calibration of the corresponding global modes, and (iii) acoustic measurements by a microphone, synchronised with PIV, used to validate the noise calculations from the model.
\begin{figure}
     \centering
         \includegraphics[width=\textwidth]{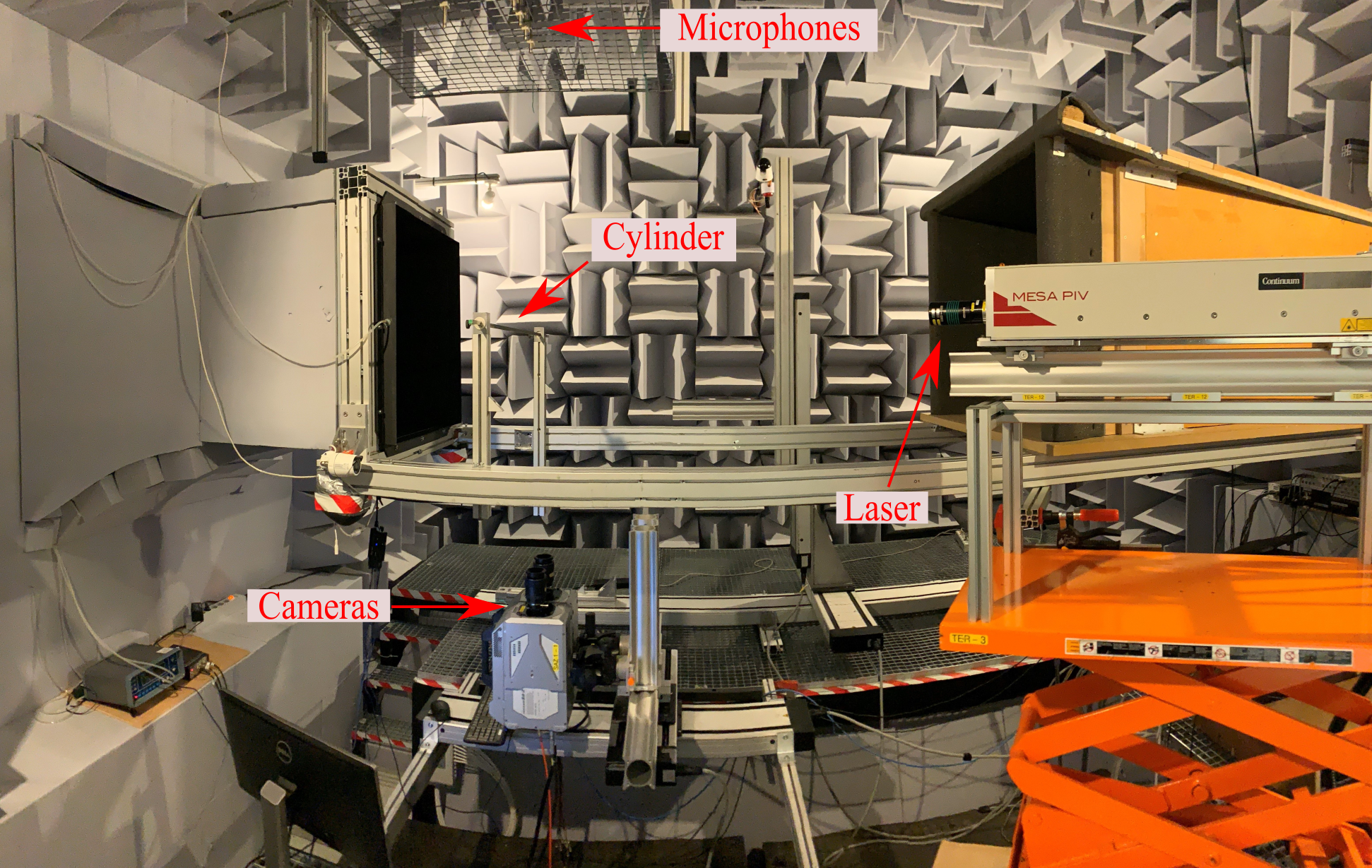}
        \caption{BETI Wind tunnel: experiments for aeroacoustics of the cylinder flows. The setup shown is for spanwise TR PIV measurements taken for global mode's amplitude calibration.}
        \label{fig:beti_general_photo}
\end{figure}

The experiments are conducted at the anechoic chamber of the wind-tunnel BETI (Bruit Environnement Transport Ingénierie) of Institut Pprime at Poitiers, France. This facility has been previously used for the location, identification, analysis and control of aeroacoustic sources \citep{zhou2020three,beausse2022}.
The setup for the present work is shown in \autoref{fig:beti_general_photo} where the flow is exiting from the convergent nozzle in the direction from left to right. The collector is placed on the right after the open test section.
The wind tunnel has a closed circuit, with an exit nozzle of section 70 cm $\times$ 70 cm, contraction factor of 10:1, and maximum velocity of $U_\infty=50$ m/s.
The walls, floor and ceiling around the open test section of the wind tunnel are covered with dihedral pieces of foam, ensuring an anechoic behaviour for frequencies above $200$ Hz.

\begin{figure}
     \centering
\begin{subfigure}[b]{0.48\textwidth}
	  \includegraphics[width=1\textwidth]{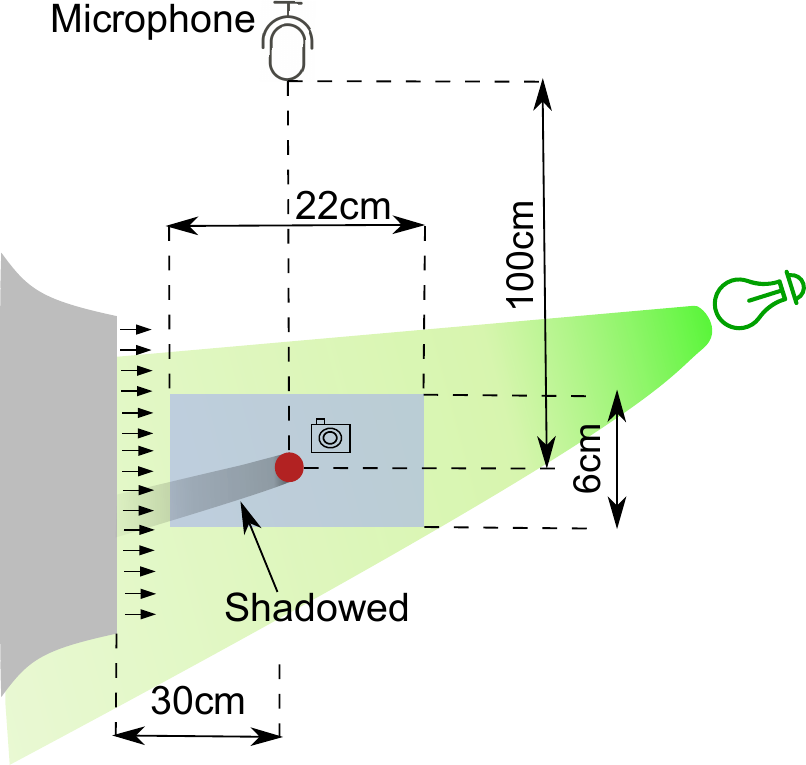}
\end{subfigure}
        \caption{Laser and camera setup for XY Mean PIV. Mean flow on which global stability analysis is done for Re=13000 flow over a cylinder (front view, not to scale).}
        \label{fig:flow_config_XY}
\end{figure}

A $d = 10$ mm cylindrical bar is placed in the test section, as shown in \autoref{fig:beti_general_photo}. 
The bar is aligned perpendicular to the incoming flow direction, positioned at the vertical centre of the test section and is located at a streamwise distance of 30 cm from the nozzle outlet.
It is mounted on two vertical stands affixed 86 cm apart which is larger than the nozzle outlet dimensions, $L=70$ cm, hence it cuts through the nozzle shear layer. 

Flow velocity was set to 20 m/s which corresponds to Reynolds numbers of 13300.
The background turbulence intensity is 1.5\%. 
The room temperature recorded during each measuring session averaged between approximately 17 and 23 degrees Celsius. Air properties were taken as air density, $\rho=1.225$ kg/m$^3$; sound speed, $c=340$ m/s, resulting in free-stream Mach number (M) of 0.06; and air kinematic viscosity, $\nu=1.5\times 10^{-5}$ m$^2$/s.

\subsection{Mean flow evaluation}\label{ssec:re13300_mean}
In this section, the mean flow evaluation is presented, which will be used for global stability analysis in the next section. 
Section \ref{sssec:xy_setup} presents the PIV measurements done to record the mean flow and Section \ref{sssec:re13300_mean} presents the post-processing of the mean flow field.

\subsubsection{Measurements} \label{sssec:xy_setup}


The incoming flow is seeded with smoke fluid and illuminated with the laser sheet in the XY plane at the mid-span ($z = 0$) via a light source kept at the top-right corner, as sketched in \autoref{fig:flow_config_XY} with the dimensions and locations of the bar and microphone. 
Note the shadowed area upstream of the cylindrical bar, which limits access to reliable velocity vector calculations in that area due to insufficient lighting.
A high-resolution camera ($4096 \times 2304$ pixels) is placed behind the laser sheet, capturing images for the spatial domains ($-11.5<x/d<0.5, -2<y/d<4$ and $-0.5<x/d<11.5, -2<y/d<4$ corresponding to the two positions for the camera). 
The time-averaged mean flows from these spatial domains were merged to have a unified domain of $-11.5<x/d<11.5, -2<y/d<4$ with spatial resolution of $\Delta x = \Delta y = 0.0235 d$. 
The velocity fields are obtained from PIV camera images by employing \enquote{\large D\footnotesize A\large V\footnotesize IS \normalsize 10} software by \large L\footnotesize A\large V\footnotesize ISION \normalsize.
A total of $6,000$ time-steps were recorded with a sampling frequency of 50 Hz corresponding to 120 s of acquisition time.

%

\begin{figure}
     \centering
\begin{subfigure}[b]{0.48\textwidth}
         \centering
	\includegraphics[width=1\textwidth]{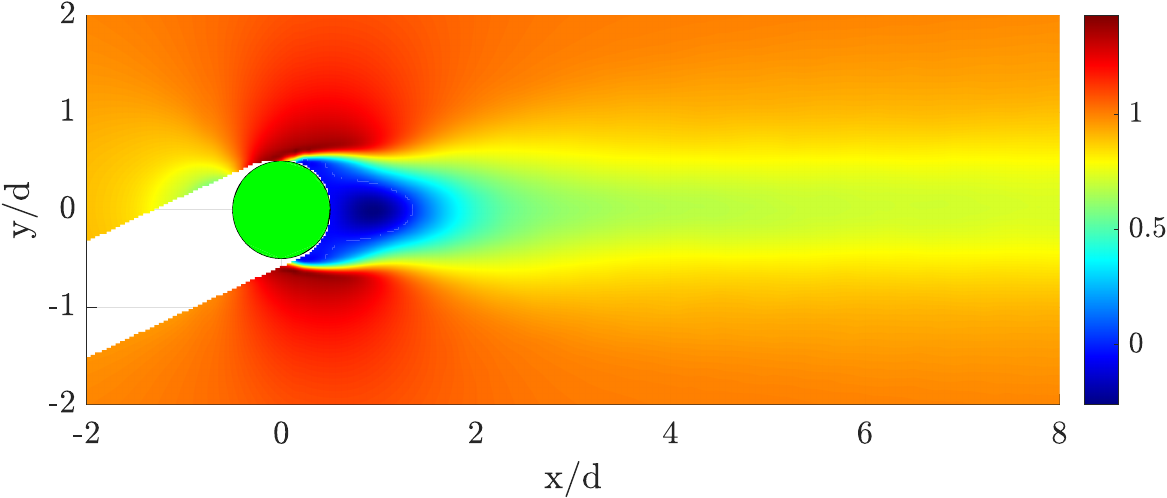}
        \caption{Time-averaged mean flow field ($\overline{U}/U_{\infty}$)}
\end{subfigure}
     \hfill
\begin{subfigure}[b]{0.48\textwidth}
         \centering
	\includegraphics[width=1\textwidth]{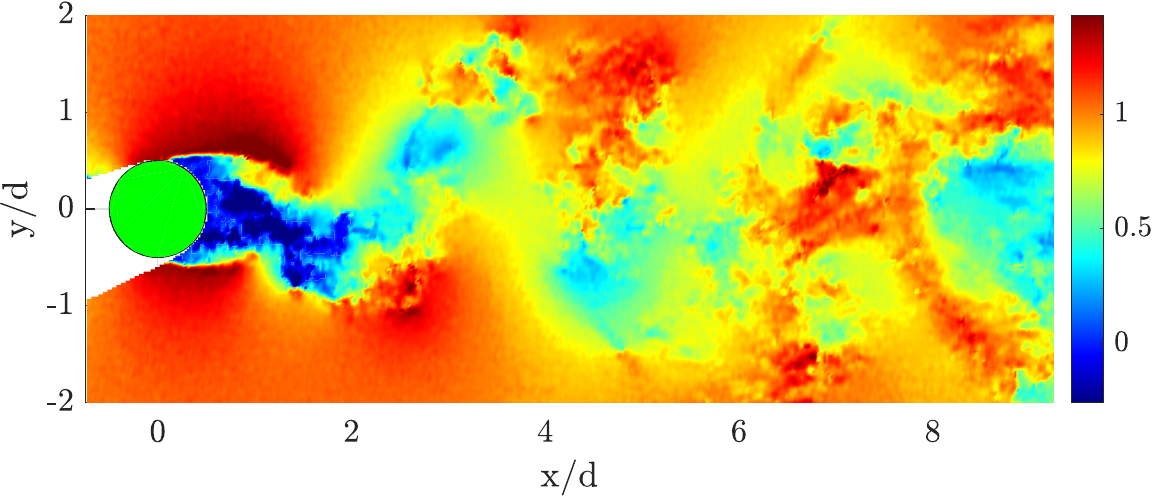}
        \caption{Velocity field at a time-instant ($U/U_{\infty}$)}
\end{subfigure}
     \hfill
        \caption{Streamwise velocity field for Re=13300 flow at $z=0$}
        \label{fig:XY_mean_timeInstant}
\end{figure}

The mean and instantaneous streamwise velocity, $\overline{U}$ and $U$ respectively, in the XY plane are shown in  \autoref{fig:XY_mean_timeInstant}. We see the expected flow separation, \autoref{fig:XY_mean_timeInstant}(a), due to an adverse pressure gradient which is followed by a recirculation region in the wake.
Observing the instantaneous velocity field snapshots, \autoref{fig:XY_mean_timeInstant}(b), we see a vortex-shedding structure in the wake of the cylinders, 
and a broad range of turbulence scales.

\begin{figure}
     \centering
\begin{subfigure}[b]{0.48\textwidth}
         \centering
	\includegraphics[width=1\textwidth]{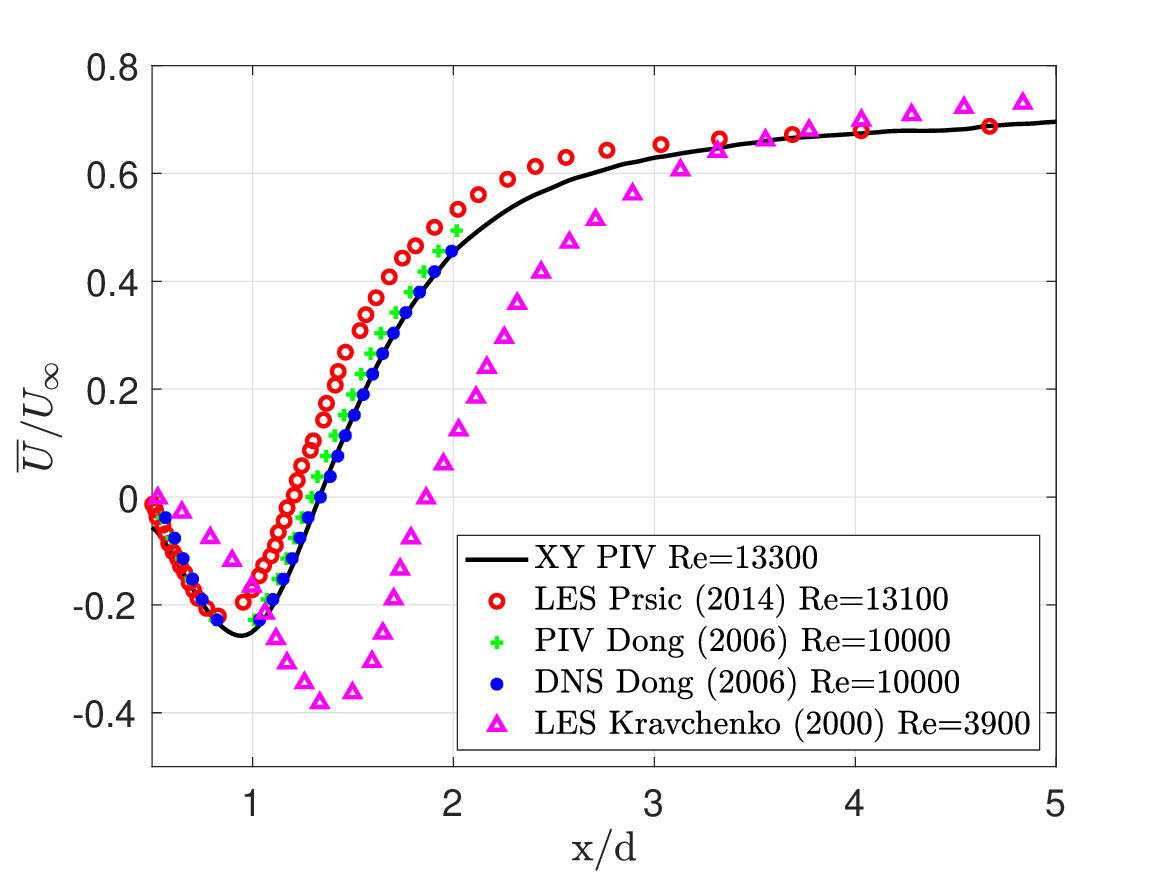}
        \caption{$x-$profile for $\overline{U}/U_{\infty}$ at $y=0$.}
\end{subfigure}
     \hfill
\begin{subfigure}[b]{0.48\textwidth}
         \centering
	\includegraphics[width=1\textwidth]{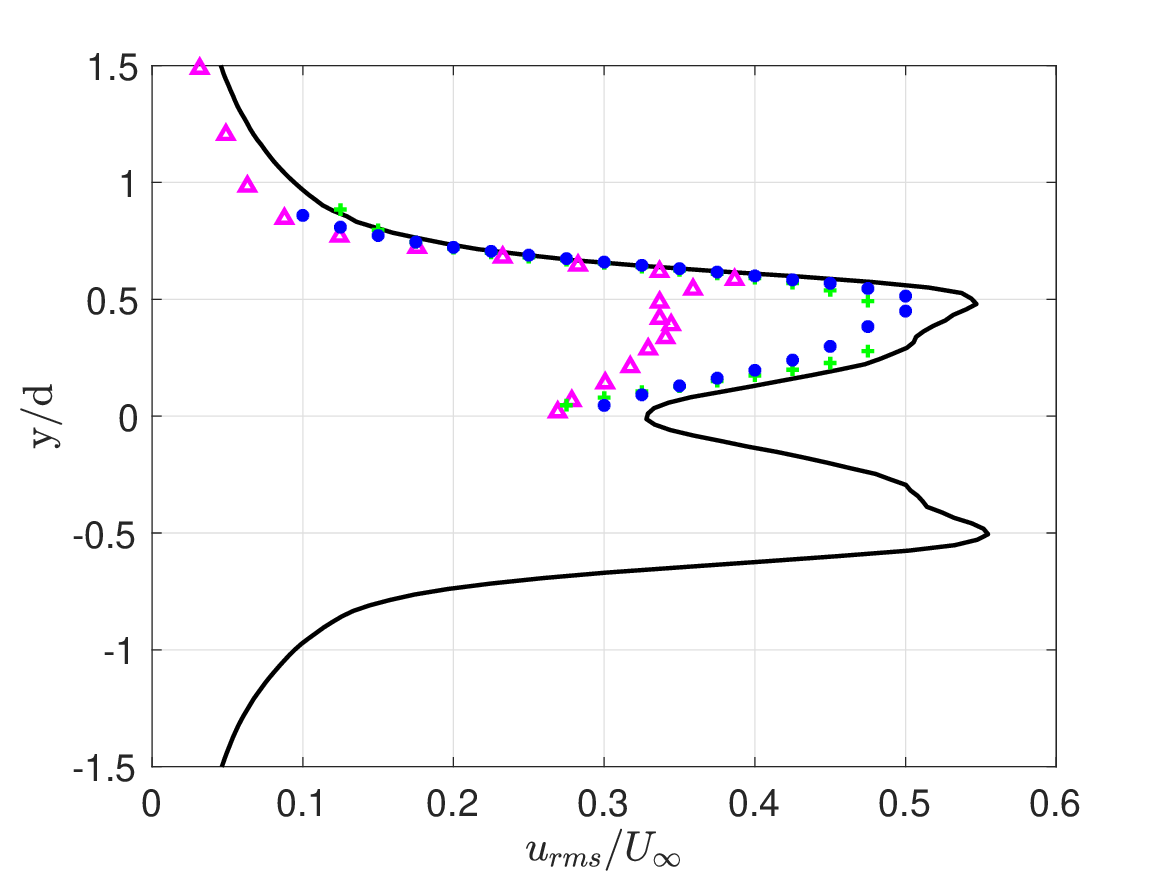}
        \caption{$y-$profile for $u_{rms}/U_{\infty}$ at $x/d=1$.}
\end{subfigure}
     \hfill
        \caption{XY Mean PIV at $z=0$: Time-averaged mean and R.M.S. for streamwise velocity}
        \label{meanU_XY_profiles}
\end{figure}

Measurements are compared with those available in the existing literature in \autoref{meanU_XY_profiles}(a), where the streamwise variation of the mean streamwise velocity at the centerline ($y=0$) is presented. 
The data is compared with Large Eddy Simulation (LES) results at Re=3900 by \citet{kravchenko2000numerical}, Particle Image Velocimetry (PIV) and Direct Numerical Simulation (DNS) results at Re=10000 by \citet{dong2006combined}, and LES results at Re=13100 by \citet{prsic2014large}. 
The presented $x$-profiles show characteristic behaviours of cylinder flow, including the distinctive minimum mean-velocity position ($\overline{U}_{min}/U_{\infty}=-0.26$ at $x/d=0.95$) and the subsequent stagnation point ($x/d=1.3$), marking the extent of the recirculation region.
A very close match is seen with the PIV and DNS results from \citet{dong2006combined} as well as a reasonable match with LES results by \citet{prsic2014large}. 

Figure \ref{meanU_XY_profiles}(b) presents the $y$-profile of the root-mean-square of streamwise velocity fluctuations at $x/d=1$. This profile shows peaks approximately at $y/d \approx 0.5$ (the centre of the shear layer), with subsequent decay as one moves away from the $y=0$ axis. 
The profile appears symmetrical across the $y=0$ axis.
A close match with PIV and DNS results from Dong et al. \cite{dong2006combined} is evidenced.
The differences with LES results by \citet{kravchenko2000numerical} are attributed to the much lower Reynolds number for the flow in their study.

\subsubsection{Post-processing}
\label{sssec:re13300_mean}
A processing of the mean flow is required as the mean flow data is unavailable in the shadowed region, and the flow velocities very close to the cylinder’s surface were contaminated by high reflection at the cylinder’s surface. The processing actions that are taken to overtake these issues are as follows:
\begin{enumerate}
\item Using the mean flow symmetry condition around the centerline, $y=0$ axis, the mean flow field in the $+y$ region is mapped symmetrically to the $-y$ region. 

\item Solution for the potential flow around the cylinder is then used to fill the remaining missing part. The potential flow $U^\phi$ around a cylinder is characterized by the velocity components given as
\begin{equation}
\overline{U}_r^\phi=\left[ 1- \frac{d^2}{4 r^2} \right]U_{\infty} \cos \theta \hspace{3mm} \mathrm{and} \hspace{3mm} 
\overline{U}_{\theta}^\phi=-\left[ 1+ \frac{d^2}{4 r^2} \right]U_{\infty} \sin \theta 
\end{equation}
where $\overline{U}_r^\phi$ and $\overline{U}_{\theta}^\phi$ are in polar coordinates, and $r=\sqrt{x^2+y^2}, \theta={\tan^{-1}(y/x)}$.

\item The mean flow is then interpolated on the unstructured grid shown in \autoref{fig:re150 domain}, the same as for Re=150 case in Section \ref{sec:results_re150}.

\item At the cylinder surface, the boundary condition (zero velocity) is forced i.e. $U=V=0$. 
\item The smoothing action is then applied to avoid the irregularities in the mean flow field that may lead to highly unstable unphysical modes while performing the global stability analysis. The smoothing utility available in the Nektar++ code package was used for this action that performs a Helmholtz smoothing projection of the form,
\begin{equation}
\left(  \nabla^2 - \left( \frac{2\pi}{L_0} \right)^2  \right) \hat{u}^{new} = - \left( \frac{2\pi}{L_0} \right)^2   \hat{u}^{orig}
\end{equation}
which can be interpreted in a Fourier sense as smoothing the original coefficients using a low pass filter of the form
\begin{equation}
\hat{u}^{\mathrm{new}}_k= \frac{1}{\left( 1+k^2/K_0^2  \right)}  \hat{u}^{\mathrm{orig}}_k \textrm{ where } K_0=2\pi/L_0
\end{equation}
where the length scale, $L_0=0.6d$ was used, below which the coefficient values are halved or more. For more details, please refer to the Nektar++ user guide.
\end{enumerate}
\begin{figure}
     \centering
\begin{subfigure}[b]{0.48\textwidth}
	\includegraphics[width=1\textwidth]{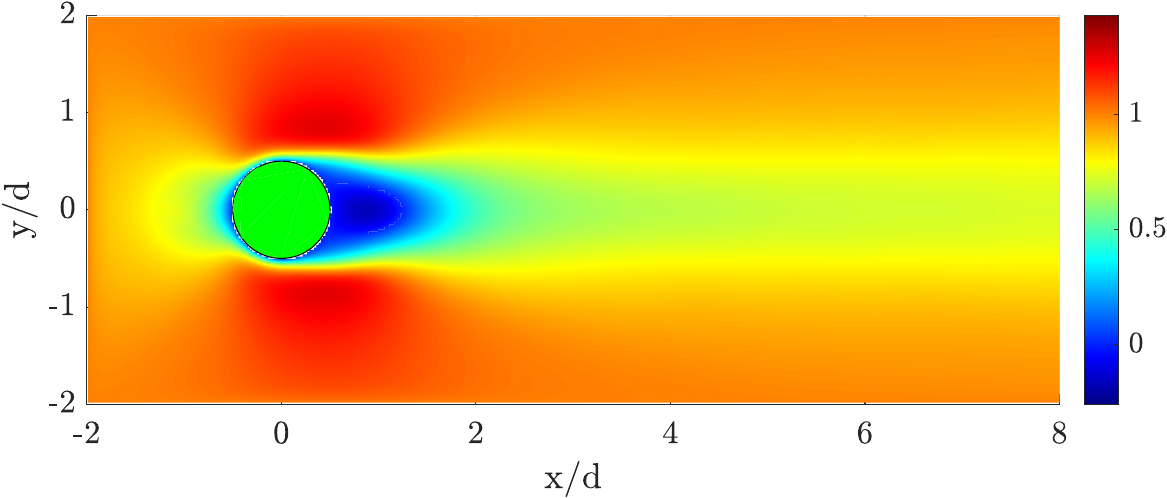}
\end{subfigure}
     \hfill
\begin{subfigure}[b]{0.48\textwidth}
	\includegraphics[width=1\textwidth]{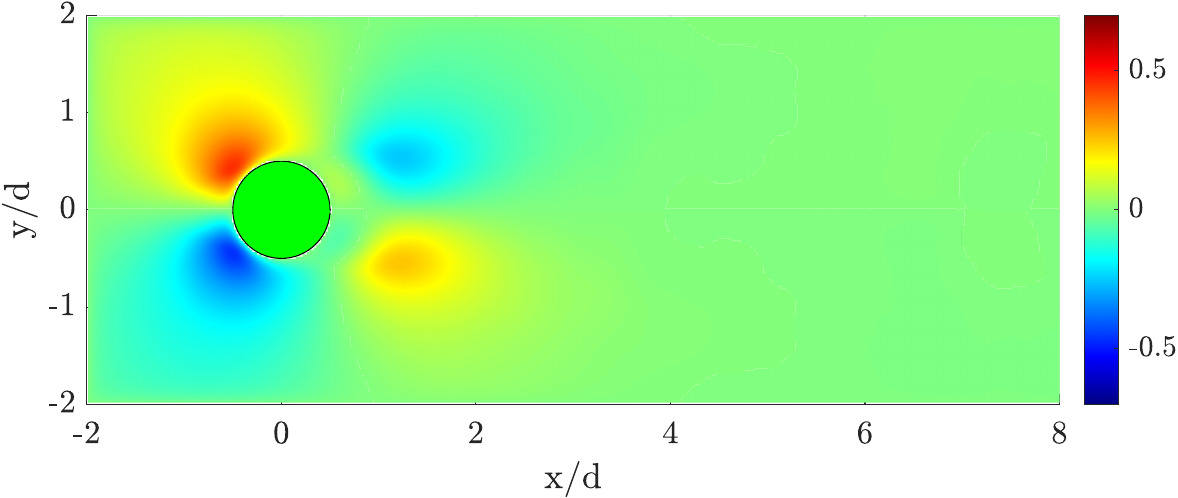}
\end{subfigure}
        \caption{Symmetrical, extrapolated, smooth, non-slip PIV mean flow over a circular cylinder at Re = 13300 flow: streamwise ($\overline{U}/U_{\infty}$, left) and transverse ($\overline{V}/U_{\infty}$ , right) components.}
        \label{fig:re13300 piv mean processing}
\end{figure}

The final processed mean flow is shown in \autoref{fig:re13300 piv mean processing} where streamwise and transverse velocities are shown. The regularities in the mean flow velocity gradients up to second order were assured.
This will be used for global stability analysis.


\subsection{Mean field, global stability analysis at Re=13300.} \label{sssec:gmc_re13300}
The global stability analysis is conducted for fluctuations of spanwise homogeneity type (i.e., $k=0$ spanwise Fourier mode) as a starting point, based on their a-priori higher acoustic efficiency. 
For the global stability analysis at Re=13300, the same numerical grid was used 
as shown in \autoref{fig:re150 domain} but with the polynomial expansions being of the order `13', which corresponds to 22308 grid points distributed over the domain of $-2<x/d<10, -2<y/d<2$.
Boundary conditions are kept as zero velocity fluctuations at the cylinder surface, left boundary, top boundary and bottom boundary, and outflow at the right boundary. 
The Krylov space of $\kappa=128$ is used for global stability analysis.

\begin{figure}
     \centering
	\includegraphics[width=0.6\textwidth]{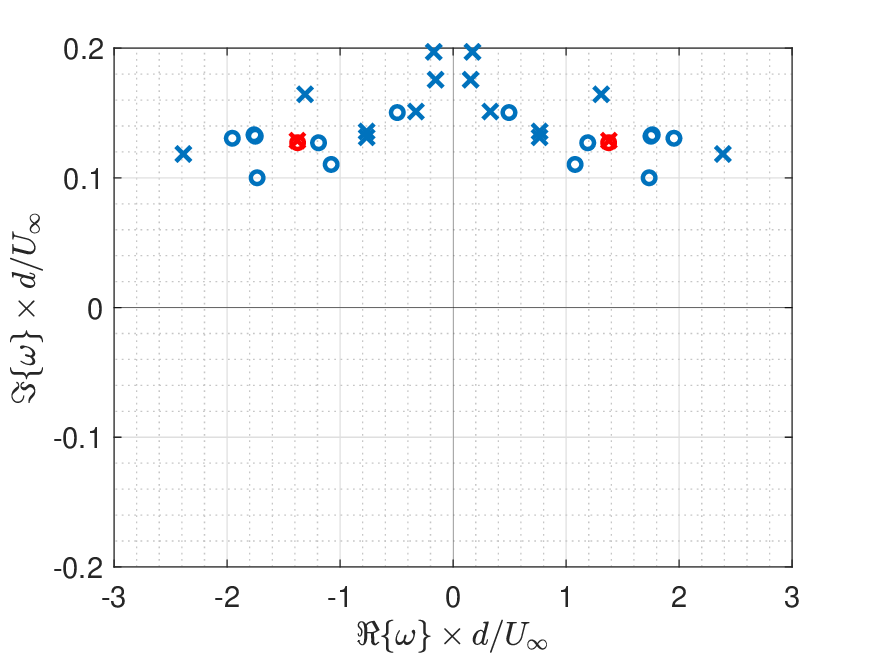}
        \caption{Eigenspectrum for the leading 16 global stability modes. Red markers represent the eigenvalues for lift fluctuation frequency mode and blue markers represent the eigenvalues for other modes. 
        Circles and crosses represent the  eigenvalues for mesh with polynomial expansions of the order  `13" in contrast to the order `14" respectively.}
        \label{fig:re13300 eigenspectrum}
\end{figure}

Among the leading eigenmodes (least decay rates), eigenvalues for the leading 16 global modes are plotted as circles in \autoref{fig:re13300 eigenspectrum}. The eigensystem is found to be globally stable.
It is furthermore observed that only one mode reaches convergence, suggesting that the remainder of the modes might be artifacts of numerical inaccuracies. Indeed, conducting the global stability analysis on a refined mesh utilizing polynomial expansions of order `14' (which corresponds to 25872 grid points in the same computational domain) returns comparative eigenspectra shown as cross symbols in \autoref{fig:re13300 eigenspectrum}: only the eigenvalue corresponding to the lift fluctuation (LF) mode remains unchanged with the grid density enhancement while the eigenvalues for the other modes exhibit significant variations.

The global mode corresponding to LF has eigenvalues
\begin{equation}
\omega \times d/U_{\infty}= \pm 1.38 + 0.13 i,
\end{equation}
which leads to St=0.220. This is close to, though slightly higher than, the dominant fluctuation frequency in the flow field  St=0.19, as reported in Section \ref{sssec:xz_setup}.

\begin{figure}
     \centering
\begin{subfigure}[b]{0.48\textwidth}
         \centering
	\includegraphics[width=1\textwidth]{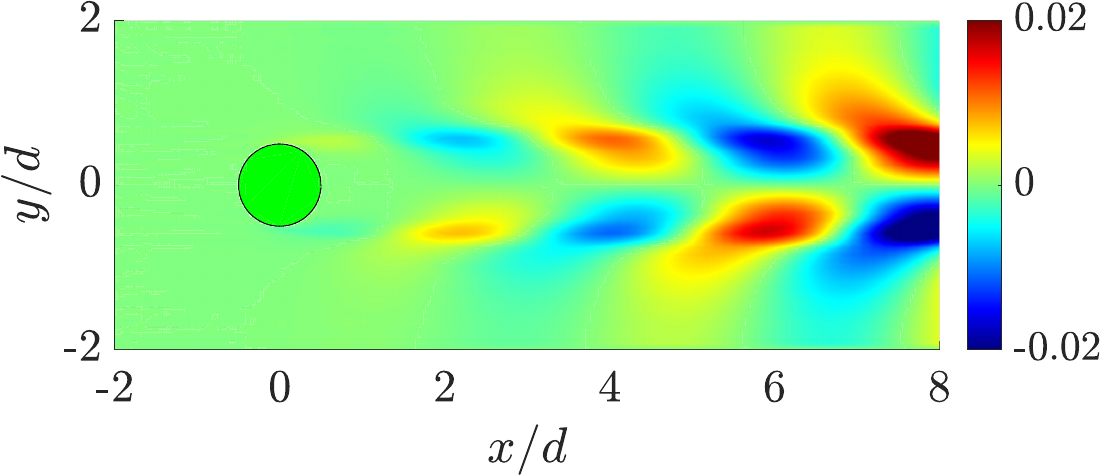}
        \caption{$	\Re\{ \tilde{u} \}/U_{\infty}$, streamwise velocity fluctuations}
\end{subfigure}
     \hfill
\begin{subfigure}[b]{0.48\textwidth}
         \centering
	\includegraphics[width=1\textwidth]{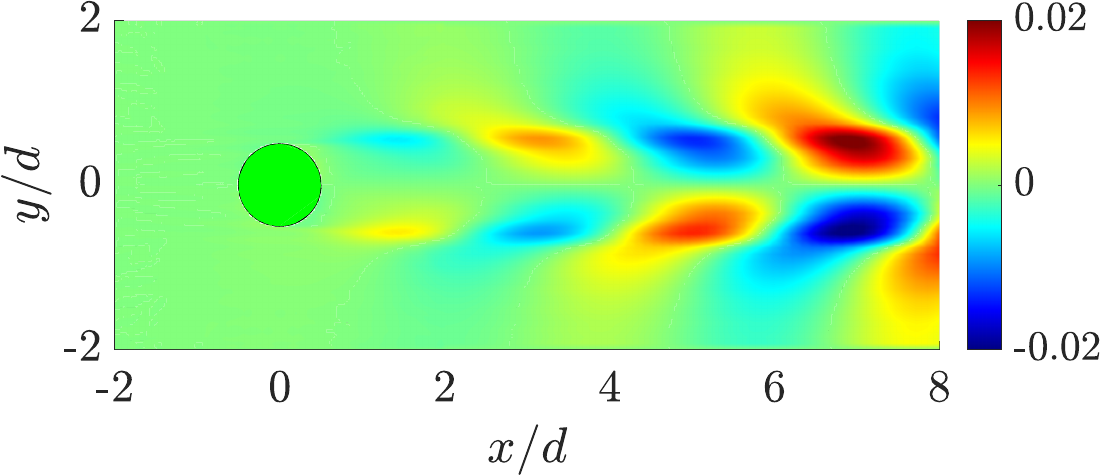}
        \caption{$	\Im\{ \tilde{u} \}/U_{\infty}$,  streamwise velocity fluctuations}
\end{subfigure}
     \vfill
\begin{subfigure}[b]{0.48\textwidth}
	\includegraphics[width=1\textwidth]{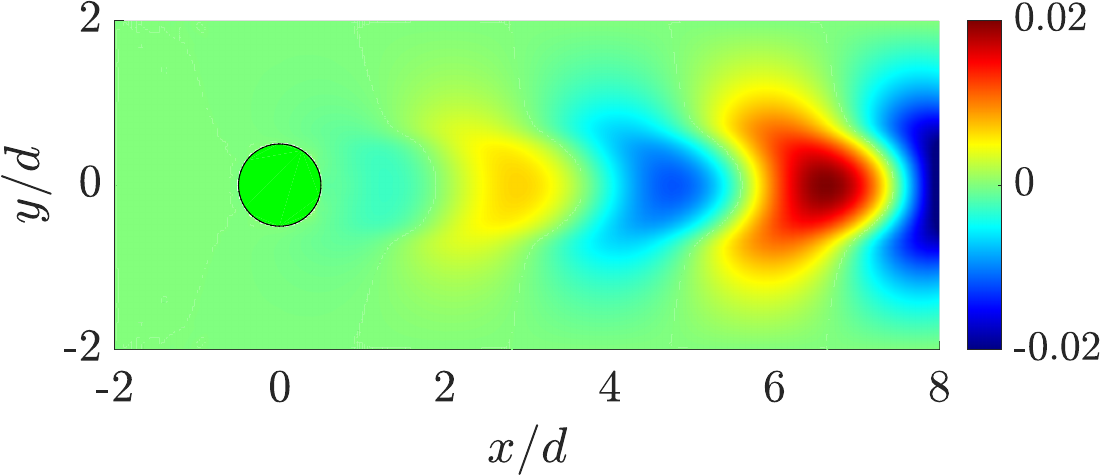}
        \caption{$	\Re\{ \tilde{v} \}/U_{\infty}$,  transverse velocity fluctuations}
\end{subfigure}
	\hfill
\begin{subfigure}[b]{0.48\textwidth}
	\includegraphics[width=1\textwidth]{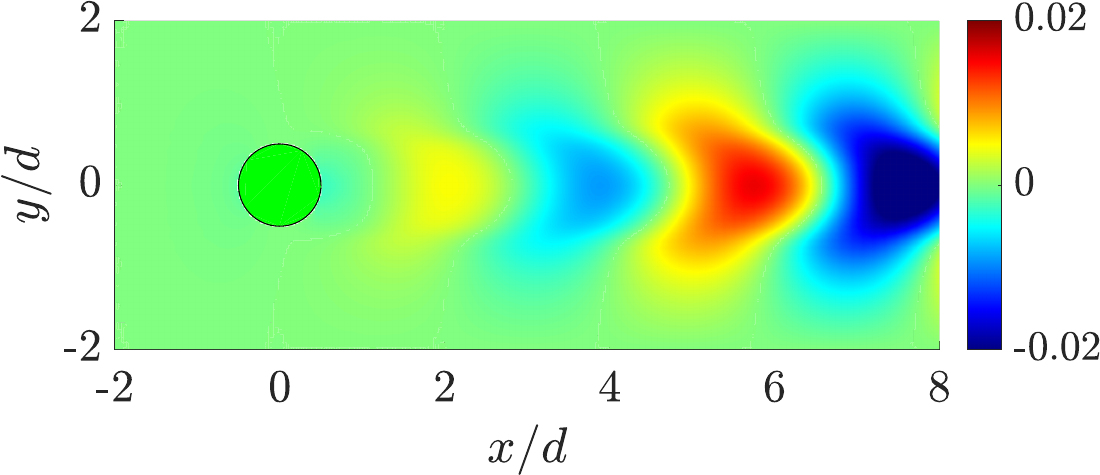}
        \caption{$	\Im\{ \tilde{v} \}/U_{\infty}$,  transverse velocity fluctuations}
\end{subfigure}
	\vfill
\begin{subfigure}[b]{0.48\textwidth}
	\includegraphics[width=1\textwidth]{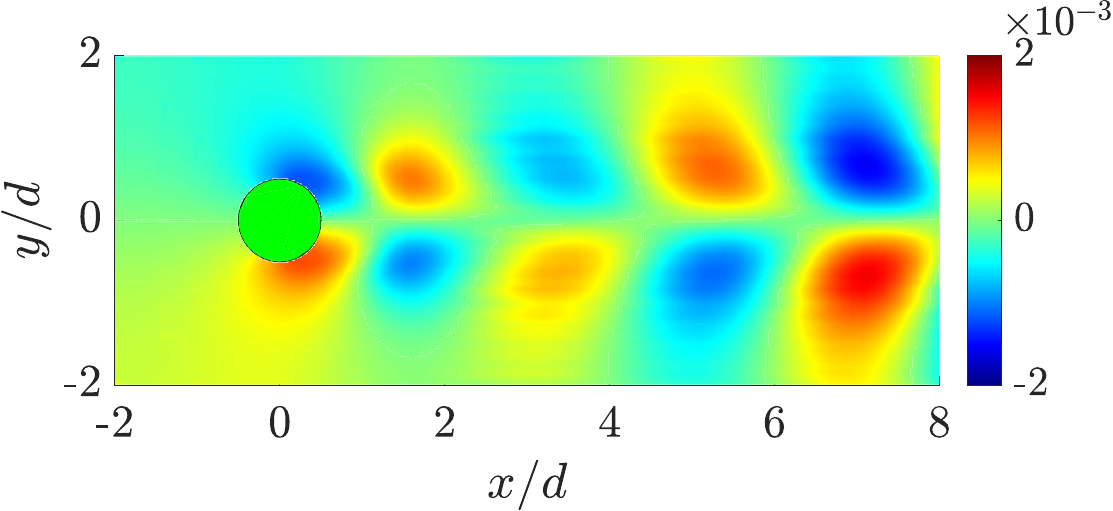}
        \caption{$	\Re\{ \tilde{p} \}/\rho U_{\infty}^2$,  pressure fluctuations}
\end{subfigure}
	\hfill
\begin{subfigure}[b]{0.48\textwidth}
	\includegraphics[width=1\textwidth]{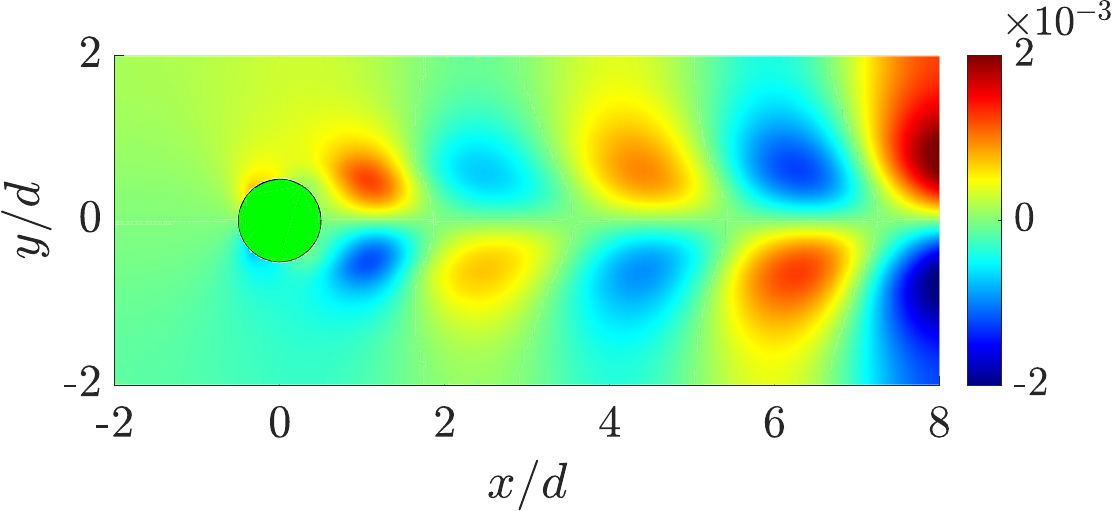}
        \caption{$	\Im\{ \tilde{p} \}/\rho U_{\infty}^2$,  pressure fluctuations}
\end{subfigure}
        \caption{Re=13300 flow over a circular cylinder: Leading global mode spatial structures}
        \label{fig:re13300 eigenstructures}
\end{figure}

The spatial structures corresponding to the wake, leading global mode are presented in \autoref{fig:re13300 eigenstructures}.
As compared to the leading Re=150 global mode, the structures are more and more pulled in the downstream direction as we move closer to the $x-$axis. 
However, it was found that, in contrast to the leading Re=150 global mode structures which grow and then decay in the downstream direction after reaching a saturation point, the leading Re=13300 global mode structures do not tend to saturate in the available domain ($x/d\leq 10$).

\subsection{Mode calibration} 
\label{sssec:re13300_calibration}
For the Re=13300 flow, amplitude calibration of the global modes corresponding to spanwise homogeneous Fourier mode is achieved by employing the Fourier-transform of the two-point measurements in the spanwise direction. 
The details of the XZ TR PIV experiment to record the velocity fluctuation time-dependent data in the streamwise-spanwise domain, and its decomposition into the frequency-spanwise wavenumber space is presented in Section \ref{sssec:xz_setup}. This data is used to calibrate the amplitudes of the leading global mode 
in Section \ref{ssssec:re13300_amplitude_calibration}.

\begin{figure}
     \centering
\begin{subfigure}[b]{0.48\textwidth}
      \centering
	\includegraphics[width=1\textwidth]{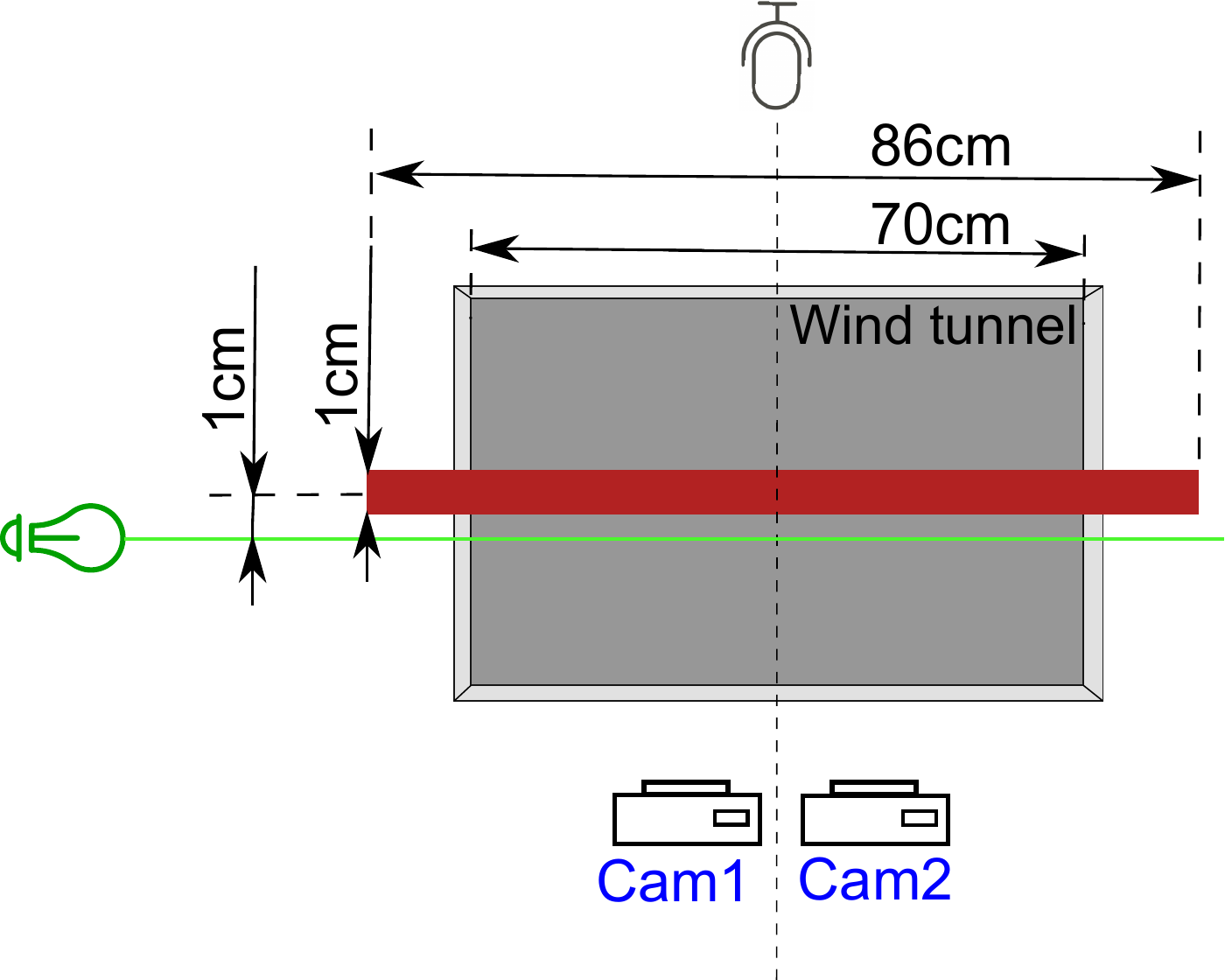}
       \caption{Side view.}
\end{subfigure}
     \hfill
\begin{subfigure}[b]{0.48\textwidth}
      \centering
	\includegraphics[width=1\textwidth]{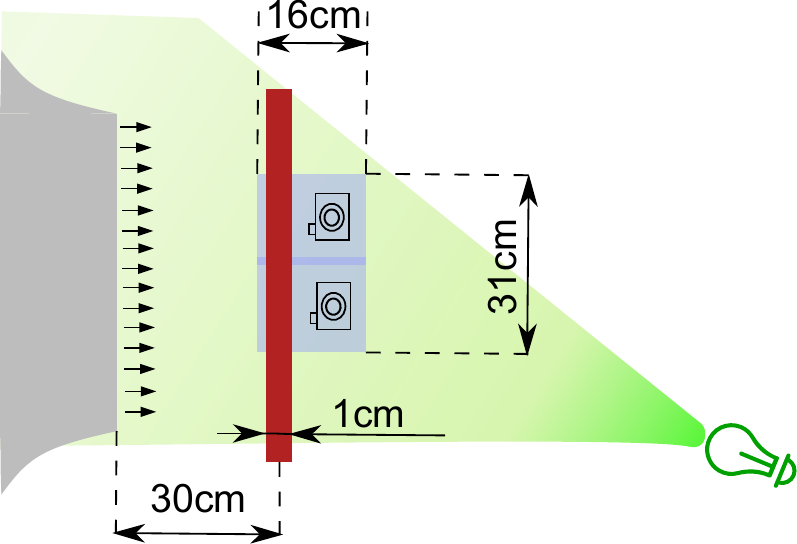}
       \caption{Top view.}
\end{subfigure}
     \hfill
        \caption{Laser and camera setup for XZ TR PIV. Spanwise TR PIV measurements for global mode's amplitude calibration. (not to scale).}
        \label{fig:flow_config}
\end{figure}

\subsubsection{Spanwise time-resolved PIV measurements} \label{sssec:xz_setup}
The experimental setup for the XZ TRPIV experimental campaign is shown in  \autoref{fig:flow_config}. 
The incoming flow is illuminated with the laser sheet in the XZ plane at $y/d=-1$. 
The decision to position the plane at $y/d=-1$ was driven by our objective to effectively capture the large-scale coherent flow fluctuations. This choice was made to ensure that the areas characterized by more intense broadband turbulence contained in the shear layer region (typically around $y/d \approx \pm 0.5$ for a circular cylinder, as illustrated in \autoref{fig:XY_mean_timeInstant} showcasing streamwise velocity fields at the mid-span XY plane), were avoided.
This strategy is similar to that of using irrotational, nearfield pressure measurements to educe the signature of coherent structures in turbulent jets 
\cite{breakey2017experimental, gudmundsson2011instability, suzuki2006instability}.

Two high-speed cameras ($1024  \times 1024 $ pixels) are placed side-to-side under the bar with an overlapping region of $-0.3<z/d<0.5$. The objective is to study the spanwise organisation of the flows in order to extract all the important time scales and length scales for the coherent structures in the flow.  Overall the two cameras give the time-dependent velocity field ($U$ and $V$) with a time resolution of $\Delta t =100\mu s$ (sampling frequency, $F_s=10$kHz) over an acquisition period of 2.18 seconds, resulting in a total of 21,842 recorded time steps.
The spatial resolution was $\Delta x = \Delta z = 0.126 d$ for the spatial domain of $-2<x/d<14, -15.5<z/d<15.7$.

%

\begin{figure}
     \centering
\begin{subfigure}[b]{0.32\textwidth}
         \centering
	\includegraphics[height=5cm]{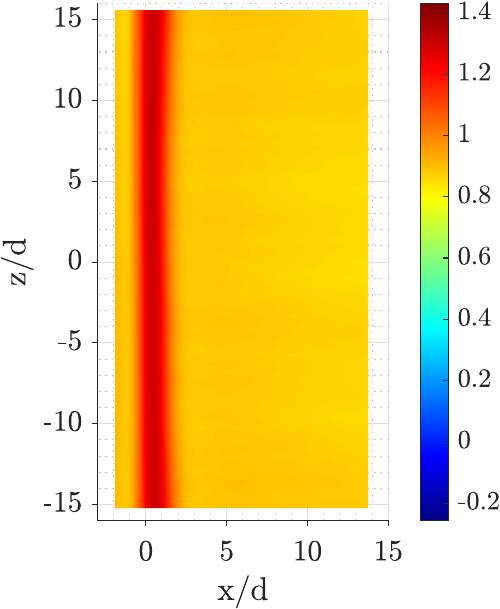}
        \caption{$\overline{U}/U_{\infty}$}
\end{subfigure}
     \hfill
\begin{subfigure}[b]{0.32\textwidth}
         \centering
	\includegraphics[height=5cm]{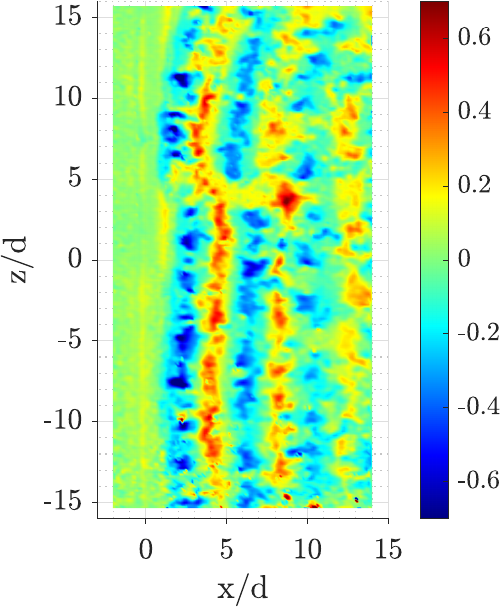}
        \caption{$u/U_{\infty}$ at a time-instant}
\end{subfigure}
     \hfill
 \begin{subfigure}[b]{0.32\textwidth}
         \centering
	\includegraphics[height=5cm]{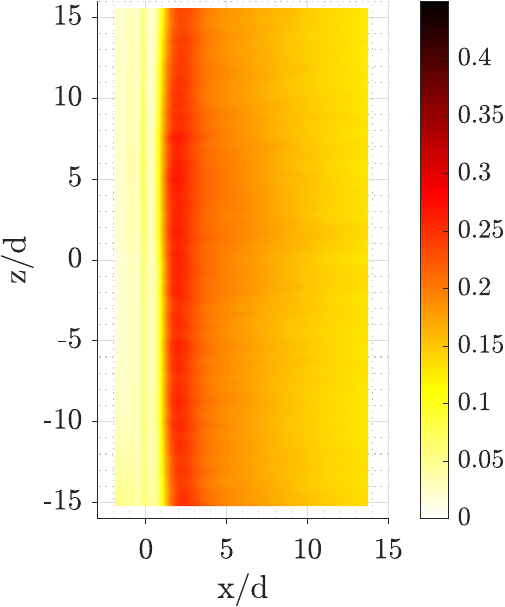}
        \caption{$u_{rms}/U_{\infty}$ }
\end{subfigure}
     \hfill
        \caption{Streamwise velocity field at $y/d=-1$ for Re=13300 flow. (a): Time-averaged mean flow field, $\overline{U}/U_{\infty}$. (b): Fluctuation field at a time-instant, $u/U_{\infty}$. (c): r.m.s. for the streamwise velocity field, $u_{rms}/U_{\infty}$}
        \label{fig:xz_mean_timeInstant}
\end{figure}

In the XZ plane at $y/d=-1$, the streamwise velocity time-averaged field ($\overline{U}$) is shown in \autoref{fig:xz_mean_timeInstant}(a). 
It looks fairly spanwise homogeneous in the considered spanwise region, $-15<z/d<15$, as was expected considering the spanwise homogeneity of the cylinders as well as the incoming flow. 
In the same XZ plane, the streamwise velocity fluctuation field ($u$) at a time-instant is shown in  \autoref{fig:xz_mean_timeInstant}(b).
We see the organized character of the vortex shedding. 
These are the structures we are interested in characterising and assessing in terms of their importance for sound radiation.
Presently, elongated spanwise structures are apparent, likely indicative of spanwise homogeneity, potentially accompanied by distinct wiggles representing higher spanwise wavenumber modes.
These coherent structures align with the alternating vortices inherent to the vortex-shedding mechanism, as also visualized in \autoref{fig:XY_mean_timeInstant} through instantaneous velocity field snapshots in the XY plane. 
We also assess the spanwise homogeneity of the streamwise velocity r.m.s. field as presented in 
 \autoref{fig:xz_mean_timeInstant}(c). 
As was the case with the mean field, the flow is, on the whole, and particularly in the mid-section, spanwise-homogeneous concerning the streamwise velocity fluctuation energy.

\begin{figure}
     \centering
	\includegraphics[height=5cm]{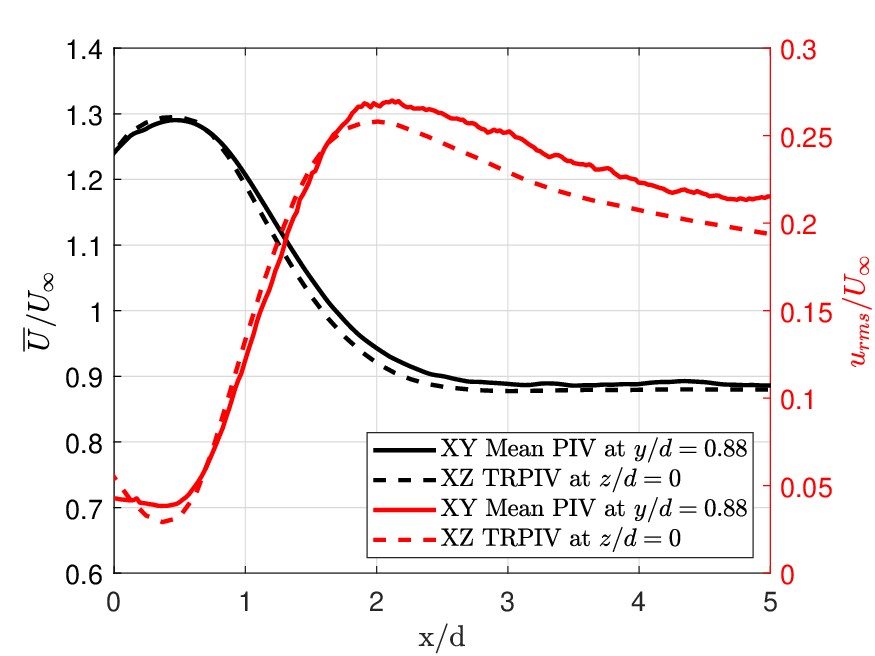}
        \caption{XY Mean PIV at $z=0$ vs XZ TR PIV at $y/d=-1$: $x-$profiles for $\overline{U}$ and $u_{rms}$}
        \label{fig:xypiv_vs_xztrpiv}
\end{figure}

The XZ TR PIV measurements are validated by comparing them against XY Mean PIV measurements at spatial locations they share, as illustrated in \autoref{fig:xypiv_vs_xztrpiv}. It is important to note that mean statistics from XY Mean PIV at $y/d=0.88$, as opposed to $y/d=1$, are presented, as they demonstrate a close alignment with XY TR PIV measurements.
This discrepancy in the $y$-location may be attributed to the misalignment of the XZ TR PIV laser sheet plane during the PIV setup. This small discrepancy of about 1 mm is generally unavoidable.
It is also minimally impactful as the spanwise coherence field has been found to be relatively insensitive to a specific $(x,y)$ location \cite{margnat2023cylinder}.

\vspace{3mm}
\noindent \textit{Fourier transform from time-space to frequency-space}
Here, streamwise velocity spanwise cross-spectral density (CSD), which is a direct measure of the acoustic source,
is evaluated, and then decomposed in the frequency-spanwise wavenumber space, which will then be used in Section \ref{ssssec:re13300_amplitude_calibration}
for the amplitude calibration of the linear global modes.

The streamwise velocity fluctuation field, $u(x,z,t)$, at $y/d=1$, is subjected to Fourier transform from time to frequency.

Following the data-processing procedure by \citet{bres2018importance}, the Welch method is used here to compute the PSD and CSD. For streamwise velocity field at each spatial location $(x,z)$, Fast Fourier transforms (FFTs) are performed on blocks of data of size $N_{fft} = 1024$, and an overlap of 75\% is imposed; i.e., block $i$ is
\begin{equation} 
Nb_i=u(x,z,1+(i-1)N_{ov}):u(x,z,N_{fft} +(i-1)N_{ov}),
\end{equation} 
where $N_{ov}=N_{fft}/4-1$.
The Hanning window $Hn$ is applied to each block prior to
application of the FFT. A Fourier-transformed block is thus
\begin{equation} 
\hat{u}(x,z,f)=\frac{\sqrt{8/3}}{N_{fft}}(FFT(Hn(t)u(x,z,t))),
\end{equation} 
where the factor $\sqrt{8/3}$ corrects for the energy loss associated with the Hanning window. The PSD of block $i$ is then computed as 
\begin{equation} 
\hat{S}_i(x,z,f)=\frac{2}{\Delta f}\hat{u}_i(x,z,f_1:\Delta f:f_{Nyq})
\hat{u}_i^*(x,z,f_1:\Delta f:f_{Nyq}),
\end{equation} 
where $\Delta f=12$Hz ($\Delta \text{St}  \approx 0.006$), $f_{Nyq}=5$kHz ($\Delta \text{St}  \approx 2.5$) and $\hat{u}_i^*$ is the complex conjugate of  $\hat{u}_i$. The block-averaged narrowband PSD is then
\begin{equation} 
\bar{S}(x,z,f)=\frac{1}{Nb} \sum_{i=1}^{Nb}\hat{S}_i(x,z,f),
\end{equation} 
where $Nb = 82$ is the total number of blocks of data.

\begin{figure}[t]
     \centering
\hfill
\begin{minipage}[b]{0.48\textwidth}
         \centering
     	\includegraphics[height=5cm]{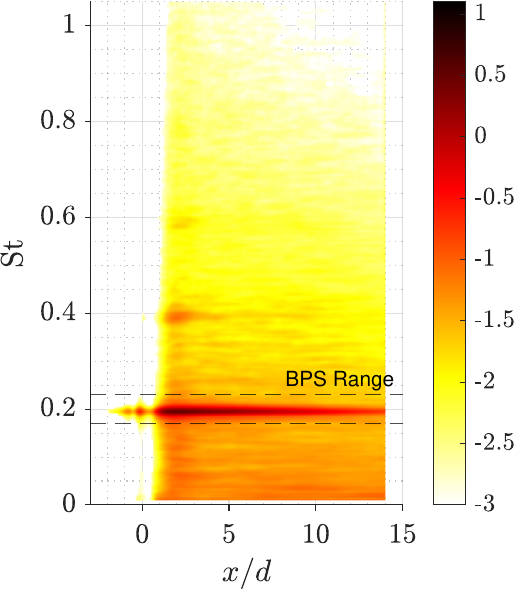}
        \caption{Streamwise velocity PSD, $log_{10}\{ \bar{S}/U_{\infty}^2 \} $, spectra at $z/d=0$, $y/d=-1$.}
        \label{fig:xz_psd}
\end{minipage}
\hfill
\begin{minipage}[b]{0.48\textwidth}
         \centering
     	\includegraphics[height=5cm]{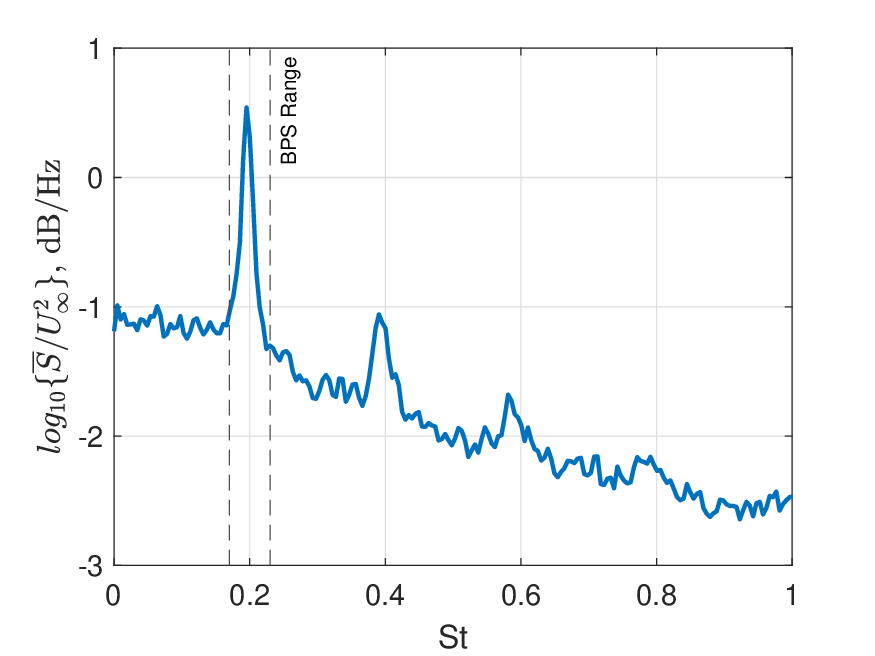}
        \caption{Streamwise velocity PSD, $log_{10}\{ \bar{S}/U_{\infty}^2 \} $, spectra at $z/d=0$, $x/d=2.01$, $y/d=-1$. Bandlimited power range, $\Delta$St$=\pm 0.03$ around St = 0.2.}
        \label{fig:xz_psd_spectra}
\end{minipage}
\vspace{1mm}
\end{figure}

At $z/d=0$ (mid-span location), \autoref{fig:xz_psd} shows the distribution of streamwise velocity PSD in the St$-x$ space.
\autoref{fig:xz_psd_spectra} shows the same at the streamwise location $x=2.01$, where peak $u_{rms}$ is located (see \autoref{fig:xz_mean_timeInstant}(c)).
We see that the majority of fluctuation energy is clustered around the LF frequency (St$\approx$ 0.2), and the same was found for other $z-$locations as well. 
Given the tonal nature of the flow and acoustic fields, our subsequent investigation focuses on this tonal frequency. 

\begin{figure}[t]
     \centering
\hfill
\begin{minipage}[b]{0.48\textwidth}
         \centering
	\includegraphics[height=5cm]{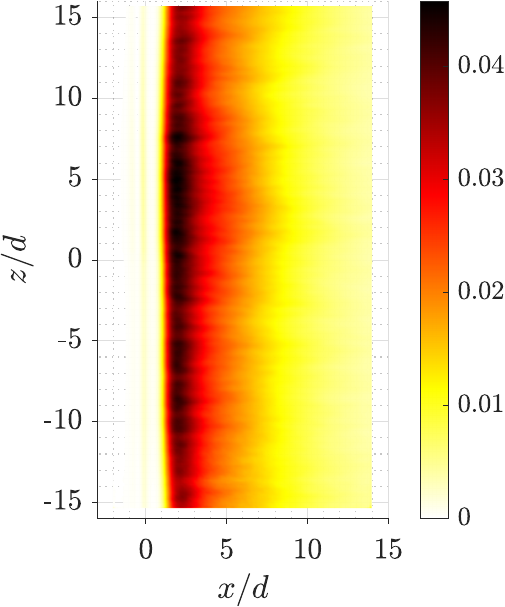}
        \caption{$BP_{PSD}/U_{\infty}^2 $, at St = 0.2 for the streamwise velocity field at $y/d=-1$.}
       \label{fig:xz_psd_Stpeak_XZcolormap}
\end{minipage}
\hfill
\begin{minipage}[b]{0.48\textwidth}
         \centering
	\includegraphics[height=5cm]{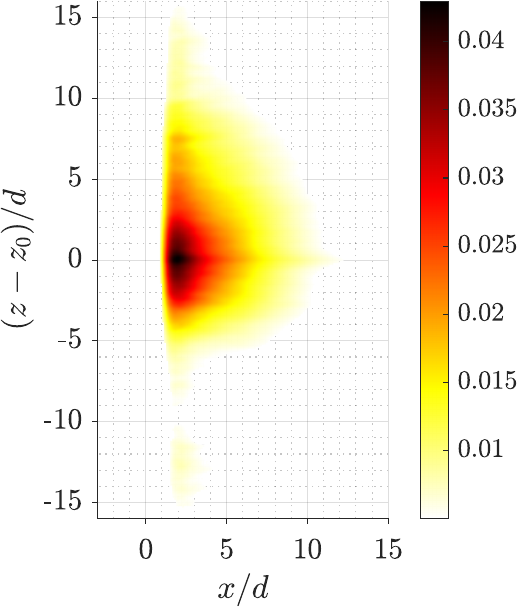}
        \caption{$BP_{CSD}/U_{\infty}^2 $, at St = 0.2 ($z_0=0$ is the reference spanwise location)}
       \label{fig:xz_csd_BP_Stpeak_XZcolormap}
\end{minipage}
\vspace{1mm}
\end{figure}

Additionally, the streamwise velocity fluctuation field here is characterized by a narrow broadband profile centred around the LF frequency as well instead of being a pure tone. 
To account for this, we consider Bandlimited Power, which in the context of streamwise velocity PSD is defined as,
\begin{equation} \label{eq:bp_psd}
BP_{PSD}(x,z) =\int_{f=f_{min}}^{f_{max}}\mid \bar{S}(x,z,f)  \mid  \hspace{1mm} df,
\end{equation} 
where $[ f_{min}, f_{max} ]$ represents the frequency range for which $BP_{PSD}$ is calculated. In the present work, it is kept $\Delta$St$=\pm 0.03$ around the LF frequency.
At LF frequency, the $BP_{PSD}$ distribution in the $x-z$ space is plotted in \autoref{fig:xz_psd_Stpeak_XZcolormap}, exhibiting good homogeneity over the PIV plane span.

Following the procedure similar to PSD computation, the spanwise CSD of block $i$ is computed as,
\begin{equation} 
\tilde{S}_i(x,z,f)=\frac{2}{\Delta f}\hat{u}_i(x,z_0,f_1:\Delta f:f_{Nyq})
\hat{u}_i^*(x,z,f_1:\Delta f:f_{Nyq}),
\end{equation} 
where $z_0$ represents the reference spanwise location relative to which the CSD is computed.
For the present work, we set $z_0=0$ i.e. the mid-span location. 
The block-averaged narrowband spanwise CSD is finally computed as,
\begin{equation} 
\tilde{S}(x,z,f)=\frac{1}{Nb} \sum_{i=1}^{Nb}\tilde{S}_i(x,z,f).
\end{equation} 
Bandlimited power for CSD is then computed as, 
\begin{equation} \label{eq:bp_csd}
BP_{CSD}(x,z)=\int_{f=f_{min}}^{f_{max}}\mid \tilde{S}(x,z,f)  \mid  \hspace{1mm} df.
\end{equation} 
At LF frequency, the distribution of $BP_{CSD}$ in the $x-z$ space is shown in  \autoref{fig:xz_csd_BP_Stpeak_XZcolormap}.
For all streamwise locations, the $BP_{CSD}$ is maximal at $z=0$ and the same as $BP_{PSD}$ at $z=0$, as per definition. 
As $\mid z \mid $ or $x$ increases, $BP_{CSD}$ decays. 

\vspace{3mm}
\noindent \textit{Fourier transform from $z-$space to spanwise wavenumber space}

The complex spanwise CSD, $\tilde{S}(x, z,f)$, is transformed in the spanwise wavenumber space through the Fourier transform as,
\begin{equation}
{\hat{\tilde{S}}}(x,k,f)= \int_{z=-L/2}^{L/2}  \tilde{S}(x, z,f) \hspace{1.3mm} e^{- i k  (z-z_0)} \hspace{1.3mm} dz,
\end{equation}
where $k$ represents the spanwise wavenumber (corresponding to the spanwise wavelength, $\lambda=2\pi/k$), ${\hat{\tilde{S}}}(x,k,f)$ signifies the Fourier transform of the $\tilde{S}(x,z,f)$ at spanwise wavenumber bin $k$, and $L=70d$ represents the cylinder spanwise length.
For the present work, $z_0=0$ has been kept as the reference spanwise location.
Note that the spanwise integration domain is limited to $[-15d, 15d]$ where data is available. This has negligible impact on the quadrature because $\tilde{S}(x, z,f)\rightarrow 0$ for $\mid z \mid > 15d$. Consequently, the following approximation is made:
\begin{equation} \label{eq:fourier_k}
{\hat{\tilde{S}}}(x,k,f) \approx
\int_{z=-15d}^{15d}  \tilde{S}(x, z,f) \hspace{1.3mm} e^{- i k  (z-z_0)} \hspace{1.3mm} dz.
\end{equation}
The computation of ${\hat{\tilde{S}}}$ by \eqref{eq:fourier_k} is performed for each frequency $f$ across all $x$ positions. Given that the available PIV domain spans $30d$ in the spanwise direction, this limitation represents the longest wavelength resolved, corresponding to $2\pi/kd=30$. On the other hand, the smallest resolved wavelengths, defined by twice the spatial resolution, result in $2\pi/kd=2\Delta z=0.25$. Notably, the zero-wavenumber term ($k = 0$) corresponds to the spanwise homogeneous component of the CSD.

Subsequently, the spectral density in the $k-$space is computed for the Bandlimited Power, $BPSD$. This methodology quantitatively evaluates how the fluctuation energy is distributed across the frequency-wavenumber space relative to the overall fluctuation energy (as depicted in  \autoref{fig:xz_mean_timeInstant}(c)). This calculation is done using the following expression:
\begin{equation} \label{eq:bp_csd_k}
BPSD_{CSD_k} (x,k)=\int_{f=f_{min}}^{f_{max}}\mid {\hat{\tilde{S}}}(x,k,f) \mid  \hspace{1mm} df.
\end{equation} 

\begin{figure}
     \centering
	\includegraphics[height=5cm]{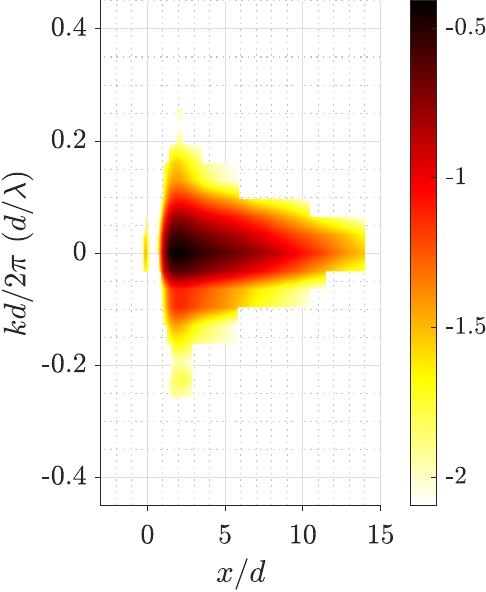}
        \caption{Spectral density in the $k-$space for the Bandlimited Power at St = 0.2, $BPSD_{CSD_k}$, $log_{10}\{ BPSD_{CSD_k}/U_{\infty}^2  \}$, for the streamwise velocity field  at $y/d=-1$. Here, $z_0=0$ is kept as reference spanwise location.}
       \label{fig:xz_CSD_Kspace_Stpeak_colormaps}
\end{figure}

 \autoref{fig:xz_CSD_Kspace_Stpeak_colormaps} illustrates the distribution of $BPSD_{CSD_k}$ in the $k-x$ space for the LF frequency.
We see that energy content diminishes significantly for wavenumbers where $\mid kd /2\pi \mid > 0.1$ (equivalently, wavelengths $\lambda < 10d$). 
This indicates that the primary energy concentration occurs within the range of wavenumbers $\mid kd/2\pi \mid < 0.1$ ($\lambda > 10d$), with particular emphasis on the dominance of the homogeneous mode ($k=0$ or $\lambda \rightarrow \infty$). 
These calculations can be used to calibrate the amplitudes of global modes.

%

\subsubsection{Mode amplitude calibration} 
\label{ssssec:re13300_amplitude_calibration}

\begin{figure}
     \centering
    \includegraphics[width=0.5\textwidth]{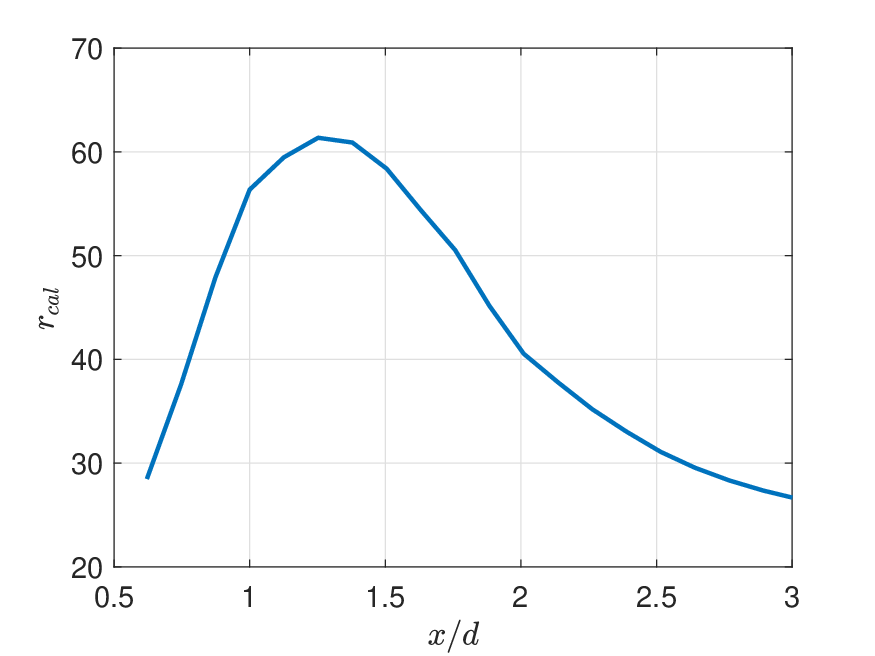}
\caption{Re=13300 flow over a circular cylinder: spanwise-homogeneous leading global mode calibration ratio, $r_{cal} (x)=\sqrt{{BP_{\textrm{PIV, }k=0}(x)}/{P_{\textrm{GM, }k=0} (x,-d)}}$. The calibration plane is at $y/d=-1$. 
} 
\label{fig:re13300_calibration}
\end{figure}

The spanwise-homogeneous component of the complex spanwise CSD, from the TR PIV measurements, is used to calibrate the leading global mode for $k=0$ spanwise wavenumber. It is expressed as
\begin{equation}\label{eq:csd_span_average}
 \tilde{S}_{k=0}(x,f)
 =\frac{\int_{z=-L/2}^{L/2}  \tilde{S}(x, z,f) \hspace{1.3mm} dz}{L}
\approx \frac{\int_{z=-15d}^{15d}  \tilde{S}(x, z,f) \hspace{1.3mm} dz}{L},
\end{equation}
which is indeed the spanwise average of the CSD. The Bandlimited power of this for the LF frequency is given as
\begin{equation} 
BP_{\textrm{PIV, }k=0} (x)=\int_{f=f_{min}}^{f_{max}}\mid  \tilde{S}_{k=0}(x,f) \mid  \hspace{1mm} df.
\end{equation} 
$BP_{\textrm{PIV, }k=0}$ is used to calibrate the leading global mode for $k=0$ spanwise wavenumber.
The power for the uncalibrated, leading global mode at the $y/d=-1$ position, is expressed as,
\begin{equation}
P_{\textrm{GM, }k=0} (x,-d)= { \mid \tilde{u} (x,-d) \mid ^2/2}.
\end{equation} 
The calibration ratio for the VS global mode amplitude is thus evaluated as,
\begin{equation}
r_{\textrm{cal}} (x)=\sqrt{\frac{BP_{\textrm{PIV, }k=0}(x)}{P_{\textrm{GM, }k=0} (x,-d)}}.
\end{equation}
The calibration ratio, $r_{\textrm{cal}}$, depends on the $x-$position in the region $0.5<x/d<3$, as shown in \autoref{fig:re13300_calibration}, where the wavemaker, the region of absolute instability, exists. Pragmatically, an average value in this spatial range is calculated 
yielding $r_{\textrm{cal}}^{\textrm{avg}}=42.5$ for the present distribution. Around this average value, a variation of $\pm 44\%$ is noticed in \autoref{fig:re13300_calibration}.

\begin{figure}
     \centering
    \includegraphics[width=0.5\textwidth]{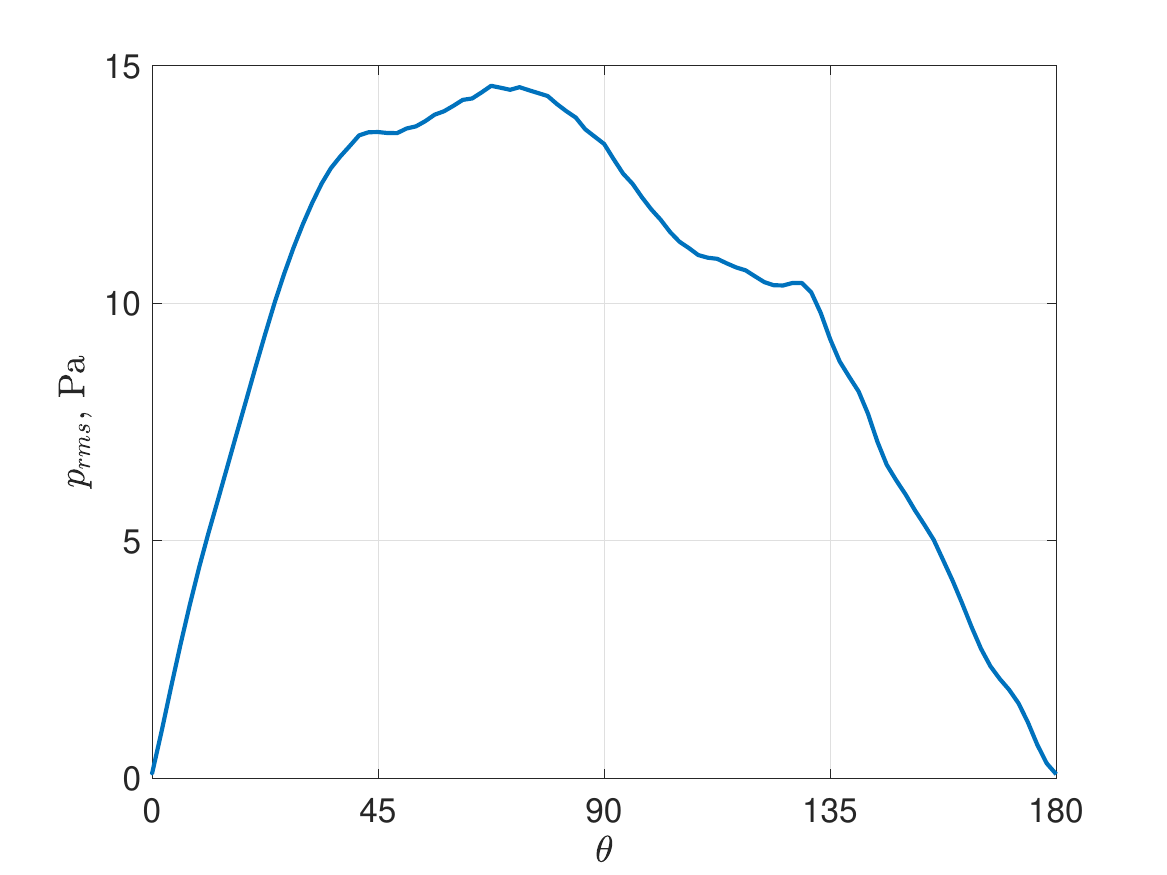}
\caption{Re=13300 flow over a circular cylinder: $p_{rms}=\mid \tilde{p} \mid/\sqrt{2} $, R.M.S. pressure fluctuations of the calibrated spanwise-homogeneous leading global mode at the cylinder surface. Streamwise direction corresponds to $\theta=0^{\mathrm{o}}$.} 
\label{fig:prms_cal_gm}
\end{figure}

The global mode is then calibrated as
\begin{equation}
\mathbf{\widetilde{\mathbf{q}}(x,y)}\mid=r_{\textrm{cal}}^{\textrm{avg}} \times \mathbf{\widetilde{\mathbf{q}}(x,y)}.
\end{equation}
On the cylinder surface, the r.m.s. pressure fluctuation of the calibrated global mode, $p_{rms}=\mid \tilde{p} \mid/\sqrt{2} $, is shown in \autoref{fig:prms_cal_gm}. The wall distribution is vanishing in the streamwise direction and is maximum around $\theta=75^{\mathrm{o}}$, consistently with the domination of lift component on the aerodynamic force fluctuation.

\subsection{Farfield noise}
\label{ssec:re13300_noisePredictions}
The calibrated global mode is used as the acoustic source input to Curle's analogy, \autoref{eq:curle_11}, 
to calculate the sound radiations associated with it. The noise calculations for the Re=13300 flow over a circular cylinder are done here and are compared with the acoustic measurements.
Acoustic measurements are presented in  Section \ref{sssec:micro_setup} and noise calculations by global mode are compared with the measurements in Section \ref{sssec:re13000_globalModeNoise}.

\subsubsection{Farfield noise measurements} \label{sssec:micro_setup}
Farfield acoustic measurements are done via a microphone  at $x/d=0, y/d=100, z/d=0$ i.e. vertically above the cylinder's mid-span location at a distance of $100d$, which is 1.42 times the cylinder’s length (see \autoref{fig:flow_config}(a)).
Microphone sensitivity is 40 mV/Pa and it is calibrated at 1 kHz, 94 decibels. 
The pressure measurements at the microphone location corresponded to the acquisition time of 6s with a sampling frequency of 50kHz. 

\begin{figure}
     \centering
	\includegraphics[height=5cm]{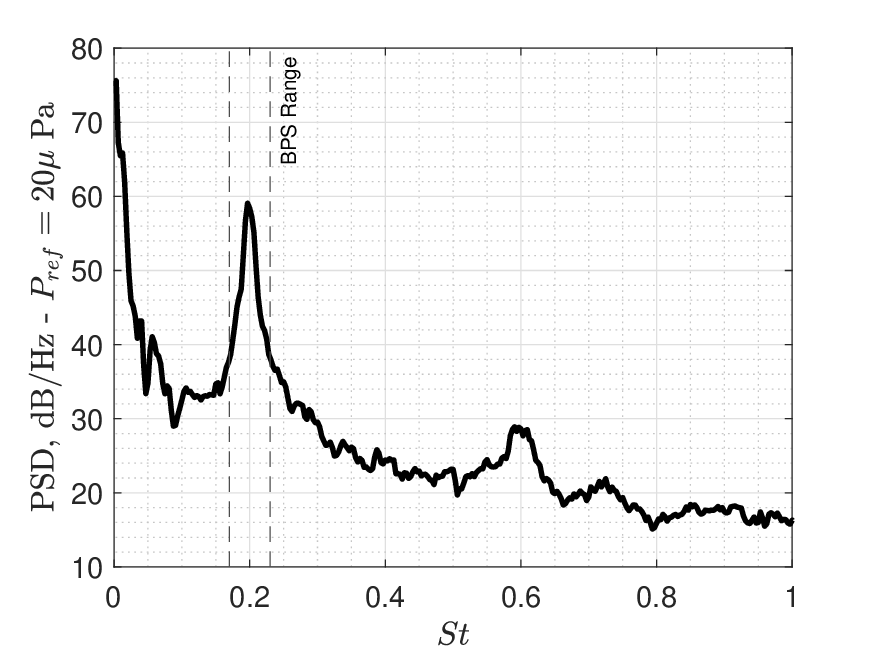}
        \caption{Microphone measurements at $x/d=0, y/d=100, z/d=0$: Noise spectra. Bandlimited power range, $\Delta$St$=\pm 0.03$ around the VS frequency.}
        \label{fig:xztrpiv_microphone_spl}
\end{figure}

The sound spectra were calculated using Welch’s method by segmenting the signal into 146 blocks (8192 time-steps in each block) with an overlap of 75 \% and Hanning window, leading to a frequency resolution, $\Delta f=6.4$Hz or $\Delta$St$=0.003$.
The resulting PSD is shown in \autoref{fig:xztrpiv_microphone_spl}. It shows that the tone appears as a discrete peak at frequency, St $\approx 0.2$ ($f=400$Hz).
Its first harmonic, St $\approx0.4$, shows negligible energy content while its second harmonic, St $\approx 0.6$ shows a local energy peak. We also see fluctuation energy accumulated for the St $<0.025$ ($f<50$Hz) which is below the anechoic frequency cutoff of the wind tunnel.

\subsubsection{Noise estimation from the global mode}
\label{sssec:re13000_globalModeNoise}
Curle's analogy has been used here as follows: the pressure fluctuations from the calibrated 2D leading global mode are homogeneously distributed (as they correspond to $k=0$ spanwise mode) along the cylinder’s span of $\pm 35d$, which is the width of the wind-tunnel jet. 

\begin{figure}
     \centering
	\includegraphics[width=1\textwidth]{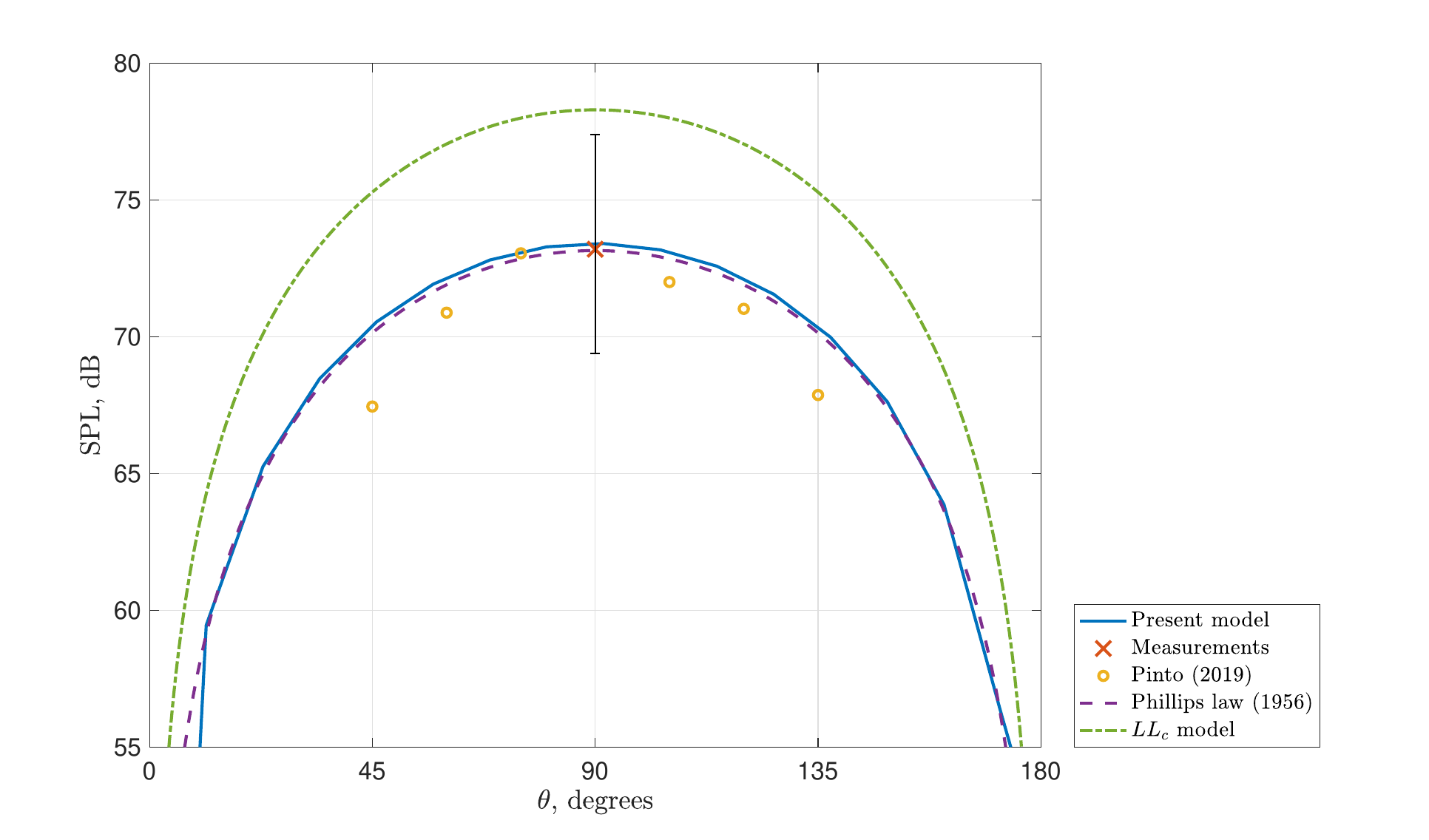}
        \caption{Re=13300 flow over a circular cylinder: SPL directivity at $r/d=100$. Streamwise direction
corresponds to $\theta=0$. 
Calibrated global mode noise calculations compared with measurements and classic models.
The $ \pm 4$dB SPL error bar (in black) at $\theta=90$ corresponds to the $ \pm 44\%$ variation in calibration ratio.}
        \label{fig:re13300_noise}
\end{figure}

Sound Pressure Level (SPL) is then evaluated as,
\begin{equation} \label{eq:spl}
\text{SPL}=20\times \log_{10} \left( \frac{p_{rms}}{p_0} \right),
\end{equation}
where $p_{rms}$ is root mean square (r.m.s) for the far-field pressure fluctuations in Pa, $p_0=20 \times 10^{-6}$Pa is the reference pressure, and SPL is the sound pressure level in dB.
The SPL noise directivity along a circular arc of radius $r/d=100$ centred at the cylinder centre is plotted in \autoref{fig:re13300_noise}, exhibiting a dipole nature.
The noise measurements at $x/d=0, y/d=100, z/d=0$ are also shown (with crosses).
The measured noise SPL is calculated from the BP in the range, St$=0.2\pm 0.03$, where the dominant tone was located. 
Also, acoustic measurements by Pinto \textit{et al.} \citet{pinto2019effect}, for the same flow configuration are plotted.

\autoref{fig:re13300_noise} shows that the noise estimation from the calibrated global mode is less than 1dB off from the measurements. 
\autoref{fig:re13300_noise} also presents the noise calculations by the classic model by \citet{phillips1956intnesity}, as expressed in 
\autoref{eq:phillips_model} and which incorporates an empirically assessed constant.
Finally, \autoref{fig:re13300_noise} includes noise estimation from $LL_c$ model as expressed in \autoref{eq:fujita_model}, using $L_c=10d$ \cite{margnat2023cylinder}, and sectional r.m.s. lift-coefficient, $C_l=0.44$ \cite{norberg2003fluc}. That model overestimates the experimental levels by more than 5 dB, which is beyond the uncertainty range of the global mode estimate due to the calibration ratio. 
Such overestimation may be attributable to the lift fluctuation coefficient, as it encompasses all the spanwise Fourier modes. It is worth revisiting the fact that not all of these modes exhibit the same acoustic efficiency, a factor that is not considered in the calculation of the coherence length.
Another source of error could be the fact that $LL_c$ model is derived under the assumption of the far-field observer's location.
However, in the current flow scenario, this assumption does not hold, as the noise predictions are made at distances where the observer is not in the far-field, specifically at $r/L=1/0.7=1.43$ or $r/\lambda=1/0.85=1.18$. 
The error associated with the assumption can still be quantified using the near-field correction factor proposed by Fujita \citep{fujita2010chars}, which yields a factor of $[1+(\lambda/2\pi r)^2]=[1+(0.85/((2\pi\times 1)))^2]=1.02$. 
Given that this correction factor is approximately equal to 1, errors stemming from the far-field assumption can be deemed negligible.

\section{Concluding remarks} \label{sec:conclusion}
A simplified physics-based low-order dynamic noise model has been proposed for Aeolian tones from the bluff bodies. 
The model is based on the representation of coherent structures, the structures dominating the acoustically important wake dynamics, as global modes from linear global stability analysis done on the mean flow. The noise model involves four key steps: (a) evaluation of time-averaged mean flow around the bluff body, (b) assessment of global modes corresponding to vortex shedding (VS) fluctuations through linear global stability analysis done on mean flow, (c) calibration of the global mode utilizing 2-point statistics in an XZ plane, and (d) noise calculations via Curle's analogy, incorporating the calibrated global mode as the acoustic source input. 

The noise model was implemented for flow over a circular cylinder at two Reynolds number cases: Re=150 and Re=13300. For the Re=150 scenario, global stability analysis on the mean flow, obtained through incompressible DNS, revealed a marginally stable dominant global mode associated with VS fluctuations (spanwise homogeneous mode, $k=0$). Calibration of this global mode, performed using 1-point statistics from DNS, showcased noise level and directivity in agreement with a direct far-field simulation, with discrepancies below $1$dB, underscoring the model's effectiveness. 

In the case of Re$=13300$ flow over a circular cylinder, global stability analysis was done on the mean flow at the cylinder’s mid-span, as determined during the XY Mean PIV campaign. Specifically analyzing spanwise homogeneous mode structures ($k=0$), the dominant global mode aligned with the lift fluctuation (LF) frequency and Aeolian tones in the far field. Amplitude calibration of this $k=0$ leading global mode utilized streamwise velocity fluctuation data from XZ TR PIV at $y/d=-1$, specifically by matching its amplitude with the corresponding frequency-wavenumber mode of the spanwise CSD. The calibration ratio was found to be dependent on the streamwise location of the calibration point and an average calibration ratio, determined in the region ($-0.5<x/d<3$) localising the wavemaker (the zone of absolute instability), was employed to calibrate the global mode.

Comparative analysis of noise directivity with microphone measurements displayed differences below $1$dB.
Despite the turbulent nature of the flow around a cylinder, featuring multiple spatial and temporal scales, the model's accuracy, built on global mode fluctuations of a rank-1 system (sole frequency: LF, singular spanwise wavenumber mode: $k=0$), is promising. 
The model was found to behave better than statistical dipole model based on the rms lift fluctuation coefficient and on the coherence or correlation length.
It would be interesting to assess how precisely it can account for change of the cross section shape, this being able to drastically modify the dipole intensity (for instance, 10 dB more for the square section~\cite{pinto2019effect}).

\clearpage
\backmatter

\clearpage
\begin{appendices}

\section{Leading global modes computation by Matrix free method based on Arnoldi Iteration} 
\label{sec:matrixfreemethods}
As mentioned in Section \ref{ssec:method_gsa}, we employ the matrix-free time-stepper approach, which is based on the modified Arnoldi method \citep{barkley2008direct} to compute the leading eigenvalues and eigenmodes.
In the present work, leading modes are the ones with largest growth rates or least decay rate if all modes are stable.

This matrix-free time-stepper approach involves performing unsteady NS simulation to generate a set of velocity fields at various time intervals. These time snapshots of velocity vectors are used to formulate a Krylov subspace which then utilised to create and solve a lower-order eigenmatrix, yielding leading eigenvectors and eigenvalues according to the desired criteria. Taken from \citet{bagheri2009matrix}, the governing system of equations is briefly derived here.

The discretized, linearized Navier–Stokes equations \eqref{eqn:lins} can be cast as an initial value problem:
\begin{equation}\label{eq:init_val_prob}
\dot{\mathbf{u}}(t) =  \mathcal{D}  \mathbf{u}(t)   \hspace{1cm} \mathbf{u}(0)=\mathbf{u}_0,
\end{equation}
for some initial state $\mathbf{u}_0$.

The long-time-horizon behaviour is determined by the
eigenvalues of $\mathcal{D}$:
\begin{equation}\label{eq:eigensystem_D}
\mathcal{D} = \mathcal{U}  \mathbf{\Lambda}  \mathcal{U}^{H} ,
\end{equation}
where the columns of the matrix $\mathcal{U}$ contain the global modes, the columns of $ \mathcal{U}^{H}$ contain the adjoint global modes (i.e. $ \mathcal{U}^{H}$$ \mathcal{U}=I$) and the diagonal matrix $\mathbf{\Lambda} =\textrm{diag}( \lambda_1, \lambda_2 ,  .... ,  \lambda_n) $ contains the eigenvalues of $\mathcal{D}$. 

Our analysis is based on the solution of the linearized Navier–Stokes equations that can be represented by the matrix exponential also referred to as the evolution operator.
The linear evolution of a perturbation under Eqs. \eqref{eq:init_val_prob} can be
expressed as
\begin{equation}\label{eq:u_timeoperator}
\mathbf{u}(t) =e^{\mathcal{D}t} \mathbf{u}_0=  \mathcal{E}(t) \mathbf{u}(0).
\end{equation}
The matrix exponential $\mathcal{E}(t)=e^{\mathcal{D}t}$ is the key to stability analysis, the eigensystem for which is represented as
\begin{equation}\label{eq:eigensystem_E}
\mathcal{E} (t) = \mathcal{U}  \mathbf{\Sigma}  \mathcal{U}^{H} ,
\end{equation}
where $\mathbf{\Sigma}=\textrm{exp}(\mathbf{\Lambda} t)$.
Note that the evolution operator for a fixed $t$ has the same eigenfunctions as $\mathcal{D}$. The temporal growth rate and frequency of the eigenmodes are given by
\begin{equation} \label{eq:eigen_relation1}
\Re(\lambda_j) = ln(\mid \sigma_j \mid )/t \hspace{1cm} 
\Im(\lambda_j) = arg(\sigma_j )/t ,
\end{equation}
respectively, where $\Sigma =\textrm{diag}( \sigma_1, \sigma_2 ,  .... ,  \sigma_n)$. If $\Re(\lambda_j)>0$ (or $\mid \sigma_j \mid>1$), the flow is considered linearly globally unstable.
We seek, for some arbitrary time $\Delta t$, the dominant eigenvalues and eigenmodes of the operator $\mathcal{E}(\Delta t)$, which has the same size as $\mathcal{D}$.

An iterative Arnoldi method \citep{arnoldi1951principle, lehoucq1998arpack} is applied to get the leading eigenmodes at a much lower computational cost.
Repeated actions of $\mathcal{E}(\Delta t)$ are applied to the discrete initial state $\mathbf{u}_0$ using the non-linear numerical simulation \citep{bagheri2009matrix, cantwell2015nektar++},
 allowing us to formulate a Krylov subspace given as
\begin{equation}
\mathcal{K}_d(\mathcal{E},\textbf{q}_0)= \textrm{span} \left\{ \textbf{q}_0, \mathcal{E}(\Delta t) \hspace{1mm} \textbf{q}_0, \mathcal{E}(2\Delta t) \hspace{1mm} \textbf{q}_0, .... , \mathcal{E}((d-1)\Delta t) \hspace{1mm} \textbf{q}_0 \right\}.
\end{equation}
where $\mathbf{u}_0$ is the initial guess that should contain non-zero components of the eigenmodes.

The Krylov subspace $\mathcal{K}$ is then orthonormalized with an $m$ step Arnoldi factorization yielding the unitary basis $\mathcal{V}$ on which $\mathcal{E}(\Delta t)$ can be projected as
\begin{equation}
\mathcal{E}(\Delta t) \approx \mathcal{V} \mathcal{R} \mathcal{V}^*
\end{equation}

This results in a much smaller $m\times m$ eigenvalue problem of the upper Hessenberg matrix  $ \mathcal{R}$,
\begin{equation}
\mathcal{R} \mathcal{S}=\mathrm{\Sigma}\mathcal{S},
\end{equation}
solvable using standard methods like the QR algorithm.

A set of Ritz values $\Sigma =\textrm{diag}( \sigma_1, \sigma_2 , .... , \sigma_m)$ generally converges rapidly to the eigenvalues of the system $\mathcal{E}(\Delta t)$. 
The Implicitly Restarted Arnoldi Algorithm (IRAM) \citep{sorensen1992implicit} integrated into the ARPACK software package \citep{lehoucq1998arpack} enables faster convergence even in case of a smaller size of Krylov subspace.
Finally, leading eigenvectors corresponding to the $m$ converged eigenvalues are recovered by $\mathcal{U}=\mathcal{V} \mathcal{S}$, and the eigenvalues for the original system are then recovered by \eqref{eq:eigen_relation1}.

In the present work, Nektar++ code package \citep{cantwell2015nektar++} is used as the numerical tool for the global stability calculations.
Information on how to install the libraries, solvers, and utilities is available on the webpage www.nektar.info. 
Mean flow, around which the global stability needs to be performed, and a starting fluctuation field, which could be a random field as well, are required as input files for the stability analysis code. At the end of every iteration, the leading set of eigenvalues and eigenvectors is obtained which is then used to create a starting fluctuation field for the next iteration. This is repeated until a converged set of eigenvalues and eigenvectors is reached.

\end{appendices}

\clearpage
\bibliography{references} 

\end{document}